\documentclass[useAMS,usenatbib]{mn2e}

\usepackage{graphicx}
\usepackage{amssymb}
\usepackage{amstext}
\usepackage{url}
\usepackage{amsmath}
\usepackage{setspace}
\usepackage{xtab}
\usepackage{textcomp}
\usepackage{booktabs,array}
\usepackage{epstopdf}
\usepackage{color}
\usepackage{colortbl}
\usepackage{tabularx}
\usepackage{gensymb}
\usepackage{subfigure}
\usepackage{placeins}
\usepackage{float}
\usepackage[normalem]{ulem}
\usepackage[english]{babel}
\usepackage{blindtext}

\title[Mode identification of $\gamma$ Doradus star HD\,12901]{Spectroscopic Pulsational Frequency
Identification and Mode Determination of $\gamma$ Doradus Star HD\,12901 \thanks{This paper
includes data taken at the Mount John University Observatory of the
University of Canterbury (New Zealand), the McDonald
Observatory of the University of Texas at Austin (Texas, USA), and the
European Southern Observatory at La Silla (Chile).}}
\author[E. Brunsden, K. R. Pollard, P. L. Cottrell, D. J. Wright, P. De Cat]{E. Brunsden$^{1}$\thanks{E-mail: emily.brunsden@gmail.com}, K. R.
Pollard$^{1}$, P. L. Cottrell$^{1}$, D. J. Wright$^{2}$, P. De Cat$^{3}$\\
$^{1}$Department of Physics and Astronomy, University of Canterbury, Private Bag
4800, Christchurch, New Zealand\\
$^{2}$Department of Astrophysics, University of New South Wales, Sydney, NSW 2052
Australia\\
$^{3}$Royal Observatory of Belgium, Ringlaan 3, 1180 Brussel, Belgium}
\begin{document}


\pagerange{\pageref{firstpage}--\pageref{lastpage}} \pubyear{2012}

\maketitle

\label{firstpage}

\begin{abstract}
Using multi-site spectroscopic data collected from three sites, the frequencies and pulsational modes of the $\gamma$ Doradus star HD\,12901 were
identified. A total of six frequencies in the range 1-2~d$^{-1}$ were observed, their identifications supported by multiple line-profile measurement
techniques and previously-published photometry. Five frequencies were of sufficient signal-to-noise for mode identification and all five displayed
similar three-bump standard deviation profiles which were fitted well with ($l$,$m$)=(1,1) modes. These fits had reduced $\chi^2$ values of less
than 18. We propose that this star is an excellent candidate to test models of non-radially pulsating $\gamma$ Doradus stars as a result of the
presence of
multiple (1,1) modes. 
\end{abstract}


\section{Introduction}

The identification of the geometry of a $\gamma$ Doradus type pulsation remains one of the more difficult spectroscopic fields. However, it is 
a powerful way to characterise the interior structure of a star and improve stellar pulsation models. The pulsations of $\gamma$ Doradus stars are
particularly of interest as they propagate through the deep layers of a star. The flux-blocking mechanism at the base of the surface
convective zone
\citep{2000ApJ...542L..57G} is held
accountable for the origin of the pulsations which are g-modes, where the
restoring force is gravity. 

Briefly $\gamma$ Doradus stars are slightly evolved late-A to late-F stars at the cooler end of the classical instability strip that display
high-order g-mode pulsations with frequencies on the order of 1 cycle-per-day. For a full definition see \citet{1999PASP..111..840K} and recent
reviews of the field
are in \citet{2007CoAst.150...91K} and \citet{2009AIPC.1170..455P}. Currently there are less than 100 bright bona fide
$\gamma$ Doradus stars known (see list in \citealt{2011AJ....142...39H}) and a handful of $\gamma$
Doradus/$\delta$ Scuti hybrid stars \citep{2005AJ....129.2026H,2008AandA...489.1213U}, with a further 100 $\gamma$ Doradus and 171 hybrid stars
thus far reported by the \textit{Kepler} mission \citep{2010ApJ...713L.192G,2011A&A...534A.125U}.

Now is truly the age of satellite photometry and no ground-based methods can compete with the long uninterrupted datasets that satellites such as
\textit{CoRoT} and \textit{Kepler} produce. The Fourier frequency spectra of such studies show the high-levels of precision only obtainable from space
(e.g. at least
840 frequencies found in HD\,49434 using the \textit{CoRoT} satellite; \citealt{2011AandA...525A..23C}). However, to further our
understanding of g-mode pulsations we need more data than frequencies alone. Several successful techniques using ground-based multi-colour photometry
can be employed to determine the number of nodal lines ($l$) in a star. The full mode-identification of a star, that is finding the $l$ and
also the number of nodal lines passing through the pole of a star $m$, remains the sole domain of spectroscopy. It is hoped that modelling of
the frequencies and modes of a star will allow determination of the $n$, number of shells interior to the stellar surface.  

The spectroscopic study of $\gamma$ Doradus stars relies on the collection of a large amount of high-resolution and preferably multi-site data of
sufficient signal-to-noise for classification of the
pulsations. Improvement is gained by the use of data from multiple sites to reduce daily aliasing patterns in the Fourier spectra. Such
observational campaigns take many months to years in order to collect a sufficient number of spectra.

\begin{center}
\begin{table*}\caption{Observation log of the multi-site data for HD\,12901}\label{obs2}
\begin{center}
\begin{tabular}{cccccc}
Observatory & Telescope & Spectrograph & Observation range & $\Delta$T (d) & $\#$ spectra \\
\hline
{\sevensize MJUO}  & 1.0m & {\sevensize HERCULES} & Feb 2009-Sep 2011 & 935 & 478\\
La Silla & 1.2m & {\sevensize CORALIE} & Nov 1998-Nov 2002& 1166 & 53\\
McDonald & 2.1m & {\sevensize SES} & Sep 2009-Oct 2009& 8 & 60\\
\end{tabular}
\end{center}  
\end{table*}
\end{center}

When such datasets are compiled for $\gamma$ Doradus stars, the pulsational frequencies and modes can be examined and compared to those from
photometry. There are still only a handful of $\gamma$ Doradus stars with full mode identifications and our immediate goal is to classify as many as
we can. The results of these mode identifications can be used to improve pulsational models (such as those
of \citealt{2012MNRAS.422L..43G,2003MNRAS.343..125T}) by providing information about the amplitudes of excited modes and also to start to identify
patterns within the class.

This paper focuses on the $\gamma$ Doradus star HD\,12901 outlining the observations made and reduction procedure in Section \ref{obs}. The
spectroscopic frequency analysis for the individual observatories and the combined data results are described in Section \ref{frequency} with each
method tested. Section \ref{phfreqID2} describes the reanalysis of white-light and seven-colour photometry taken of this star. The mode
identification of
the five identified frequencies follows in Section \ref{modeid2}. Finally Section \ref{disc} discusses the findings and their implications for future
work in this field.

\section{Observations and data treatment}\label{obs}

The major findings of this research are the identification of the frequencies and modes using spectra obtained at three observing
sites (Mt John University Observatory ({\sevensize MJUO}), La Silla Observatory and McDonald Observatory) and are summarised in Table~\ref{obs2}. 

Spectra were reduced according to the standard local spectrograph software then they were normalised, continuum fitted and order-merged by the authors
using a semi-automated {\sevensize MATLAB} routine. The full spectra were cross-correlated for each
site using a scaled delta function routine (\citealt{NewEntry3}, \citealt{2007CoAst.150..135W}). This produces the line
profiles used for pulsational analysis. The signal-to-noise of a cross-correlated line profile is much higher than a single
spectral line and thus pulsations are easier to extract. The line profiles for each
site were combined (scaled, shifted in velocity space and continuum fitted) to produce a single dataset of consistent line profiles. This was done
post-cross correlation to maximise the number of available lines used for each spectrum, as different spectrographs operate over different wavelength
ranges. 

The line profiles were analysed in {\sevensize FAMIAS}, a pulsational frequency mode-identification toolbox \citep{2008CoAst.157..387Z}, using
techniques applied
from \citet{2006A&A...455..235Z}. The Fourier spectra for the Pixel-by-Pixel (PbP)
technique and that of the 0th-3rd moments \citep{1986MNRAS.219..111B,2003AandA...398..687B} were used to identify frequencies. These frequencies were
then analysed to determine mode identifications using the Fourier Parameter Fit method \citep{2009AandA...497..827Z}. 

Additionally, the software package {\sevensize SIGSPEC} \citep{2007AandA...467.1353R} was used to compare the frequency selection
method of {\sevensize FAMIAS}. {\sevensize SIGSPEC} performs a Fourier analysis of a two-dimensional dataset and selects frequencies based on their
spectral
significance. The frequencies obtained with {\sevensize SIGSPEC} can be regarded as more precise than that of the highest-peak direct
selection as, after producing the Fourier frequency spectrum, the peak with the highest spectral significance is selected. Spectral significance
includes the analysis of the false-alarm probability to remove frequency peaks caused by irregular data sampling or noise in the data.

Both {\sevensize FAMIAS} and {\sevensize SIGSPEC} were used to re-analyse photometric data, originally published in
\citet{1999MNRAS.309L..19H}, \citet{2000AandA...361..201E} and
\citet{2004AandA...415.1079A}, as further insight into the frequencies of the pulsations. To identify the $l$ values of the modes, the frequency
amplitude ratio and phase differences method, based on pre-computed grids of models, were used
\citep{1979MNRAS.189..649B,1988ApandSS.140..255W,1994AandA...291..143C,2002AandA...392..151D}. This was done using the photometric analysis toolbox in
{\sevensize FAMIAS} \citep{2008CoAst.157..387Z}.

\section{Spectroscopic Frequency Analysis}\label{frequency}

In total $591$ spectra from the three sites were of sufficient quality to be analysed. The data spanned a total of 4667 days, just over 12 years.
Each dataset was analysed independently for frequencies and they were then combined. By doing this analysis we can comment on the limitations of
single-site data and
the extent to which multi-site data reduces aliasing.
 
\subsection{Frequency Identification For Each Dataset}

\subsubsection{{\sevensize MJUO}}

This is the largest single-site dataset and was analysed most extensively. First the cross-correlated line profiles were analysed
in {\sevensize FAMIAS}. The frequency list is in Table \ref{hercsig2}. The PbP technique found eight possible frequency peaks including a
one-cycle-per-day frequency ($f_{m5}$), likely an alias.  The zeroth moment
was too noisy to extract any frequencies. The first and third moment Fourier peaks each only showed one frequency clearly, $1.186$~d$^{-1}$ and
$1.184$~d$^{-1}$ respectively,
which is the same frequency given a conservative uncertainty estimate of $\pm 0.001$ (uncertainties are dealt with more formally in Section
\ref{freqres2}). The
second moment yielded $0.144$~d$^{-1}$ as the only viable frequency. Though many more frequencies may be
evident in the data, further extraction of frequencies above the noise was difficult with any measure of certainty.

The same moments were then tested in {\sevensize SIGSPEC} for further analysis. {\sevensize SIGSPEC} found 20 frequencies above a spectral
significance level of 5. It
is likely that noise in the data is causing many misidentifications rather than the possibility that these could all be real, given the noise level in
the Fourier spectra. The frequencies with spectral significance greater than 15 are reported for moments one to three in Table
\ref{hercsig2}. From the table it is clear that $f_{m2}$ to $f_{m8}$ (excluding $f_{m5}$) are viable frequencies found using multiple methods. The
addition of further
datasets from
other observation sites should improve the signal-to-noise and reduce any aliasing, particularly 1-day aliasing that occurs in the data.

\begin{center}
\begin{table*}\caption{Frequencies of HD\,12901 found in the {\sevensize MJUO}  data. Pixel-by-Pixel (PbP)
frequencies found using {\sevensize FAMIAS} and moments and significances (Sig) found using {\sevensize SIGSPEC} are listed. The amplitudes
of the PbP frequencies are scaled to the first identified amplitude. Note some frequencies have been
identified as $1-f$
aliases of the PbP frequency. The frequency $f_{m5}$ shows identifications near 1~d$^{-1}$, likely from the data sampling.}\label{hercsig2}
\begin{center}
\begin{tabular}{ccccccccc}
×  &  \multicolumn{2}{c}{PbP} & \multicolumn{6}{c}{Moments}\\
\hline
×  & Freq. & Amp. & $1^{st}$ & Sig & $2^{nd}$  & Sig & $3^{rd}$  & Sig\\
×  & (d$^{-1}$) & (rel. to $f_{m1}$) & (d$^{-1}$) &  &  (d$^{-1}$) &  & (d$^{-1}$)  & \\
\hline
$f_{m1}$ & 1.271 & 1   & 	&    &   	&    &		&		\\
$f_{m2}$ & 1.186 & 1.8 & 1.183  & 24 &  	&    &	1.183	&	23	\\
$f_{m3}$ & 1.681 & 1.5 & 0.678	& 26 &   	&    &	0.678	&	28	\\
$f_{m4}$ & 1.396 & 1.7 & 0.403	& 20 & 1.396 	& 20 &		&		\\
$f_{m5}$ & 1.001 & 1.3 & 	&    & 1.001 	& 26 &		&		\\
$f_{m6}$ & 1.560 & 1.6 & 0.460	& 16 & 1.560 	& 17 &	1.560	&	19	\\
         &       &     & 	&    &  	&    &	0.541	&	18	\\
$f_{m7}$ & 1.216 & 1.6 & 2.214	& 18 & 0.213 	& 23 &	2.214	&	20	\\
$f_{m8}$ & 0.244 & 1.1 & 0.756	& 21 &  	&    &		&		\\
$f_{m9}$ &       &     & 	&    & 0.104 	& 16 &	0.899	&	19	\\
$f_{m10}$&       &     & 	&    &  	&    &	2.352	&	20	\\
$f_{m11}$&       &     & 	&    &  	&    &	0.836	&	17	\\
\\
\end{tabular}
\end{center}  
\end{table*}
\end{center}

\subsubsection{La Silla}

The spectra taken on the {\sevensize CORALIE} spectrograph have previously been analysed and published by \citet{2004AandA...415.1079A} and
\citet{2006AandA...449..281D}. Frequencies of $1.04\pm0.28$~d$^{-1}$ and $1.30\pm0.30$~d$^{-1}$ were reported.

Although the small size of the dataset made it difficult to analyse, frequency peaks were identified at $0.005$~d$^{-1}$ and $1.26$~d$^{-1}$
in the PbP measure, $0.997$~d$^{-1}$ and $3.01$~d$^{-1}$ in the first moment and $0.997$~d$^{-1}$ and $1.999$~d$^{-1}$ in the third
moment. Peaks that occur very close to integer values are less reliable as they are
likely 1-day alias patterns from the observation times artificially amplified by data sampling effects in the window function. No
identifiable peaks occurred in the zeroth and second moments. Only the
$1.26$~d$^{-1}$ can be identified with any confidence. It is notable that a peak was apparent in
all
the above Fourier spectra at $0.23$~d$^{-1}$.

{\sevensize SIGSPEC} was better able to distinguish the 1-day aliases and identified frequencies of $0.26$~d$^{-1}$ in the zeroth
moment and
$0.23$~d$^{-1}$ in the first and third moments. These had significances of $3.3$, $3.5$ and $2.7$ respectively, which are usually regarded as too
small
to be significant.

\subsubsection{McDonald}

The McDonald data set comprised of 60 observations taken in September and October 2009. The data were considered to be particularly useful as
having time intervals that overlap some of the {\sevensize MJUO} observations, providing some independent confirmation of the line profile
variation. As a stand-alone data
set the shorter wavelength range of the spectrograph, and the larger regions of telluric lines, reduced the number of stellar lines that could be
cross-correlated from several thousand to around 100 lines. This has a significant impact on the signal-to-noise of the final line
profile. This, and the low number of observations, meant frequencies detected in this dataset alone were unconvincing. Despite this, it is noted that
frequencies near $1.38$~d$^{-1}$ and $1.24$~d$^{-1}$ are present in the PbP analysis and a peak near $1.67$~d$^{-1}$ is visible in the
first moment Fourier spectrum.

\begin{center}
\begin{figure}
\begin{center}
   \includegraphics[scale=0.20]{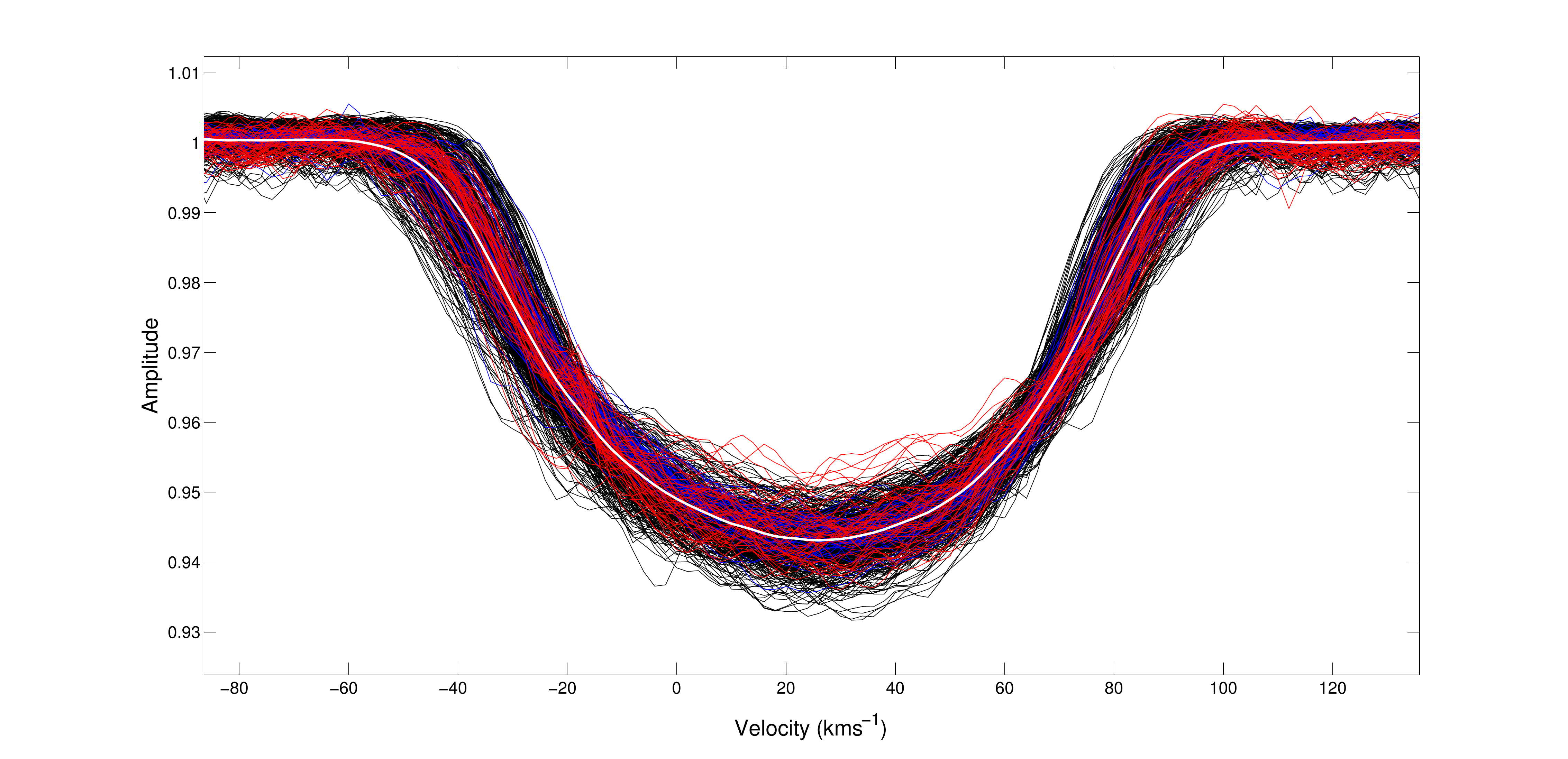}\caption{Scaled line profiles of the three datasets. McDonald (red), La Silla
(blue) and {\sevensize MJUO}  (black) observations are shown with the mean line profile (white). All datasets scale well to show a consistent line
profile. The data are provided as an electronic file (see Supporting Information).}\label{3lineprof}
\end{center}
\end{figure}
\end{center}

\subsection{Frequency Identification Of Combined Data For Each Method}

\begin{center}
\begin{figure*}
\begin{center}
   \includegraphics[scale=0.6, trim=3cm 2cm 0.1cm 0.1cm, clip=true,]{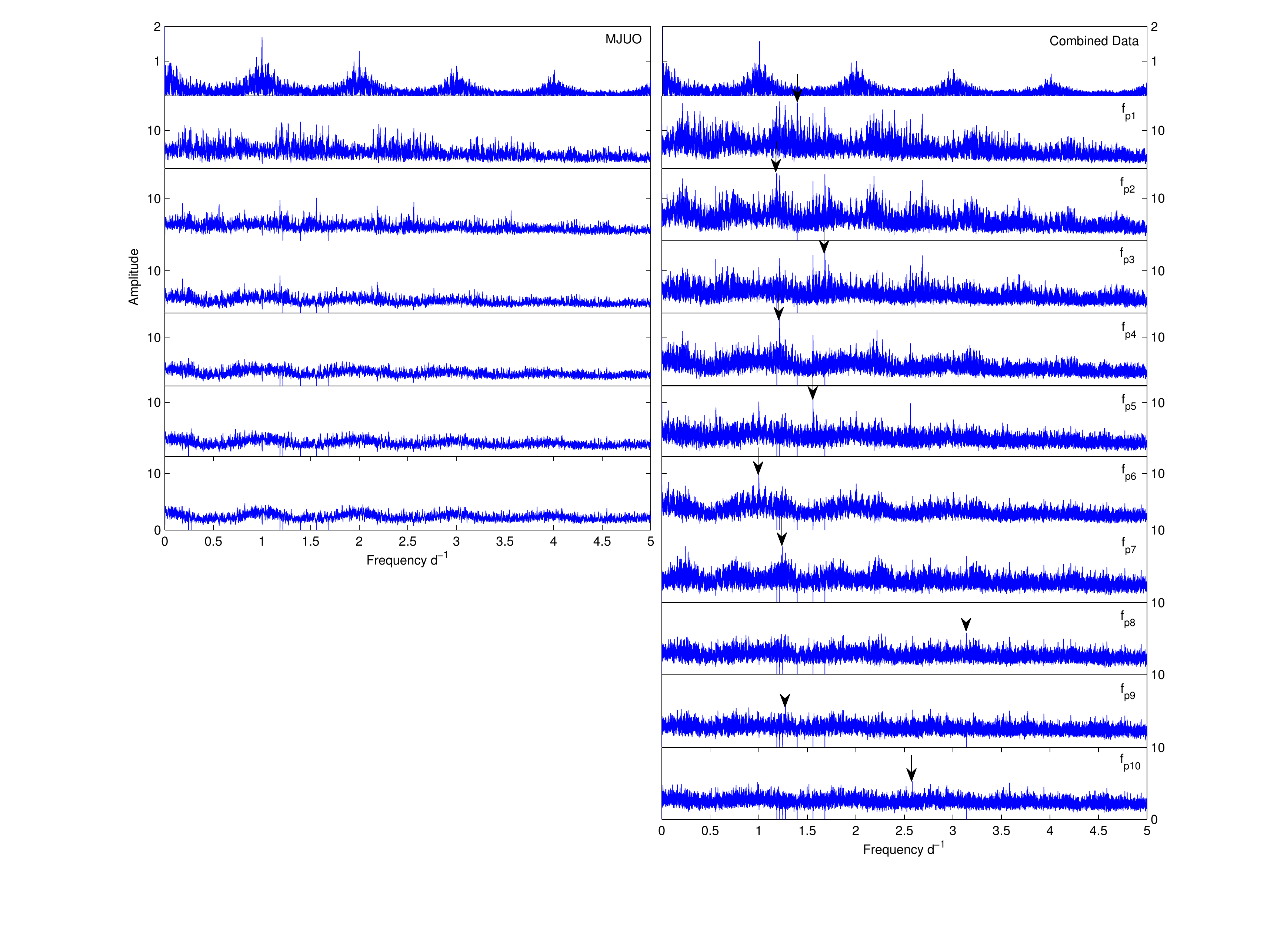}\caption{Effect of the multi-site datasets on the
Pixel-by-Pixel Fourier spectra, the MJUO data on the left and the combined data on the right. Each series shows the window function for the data at
the top then successive pre-whitening of frequencies $f_{p1}$ to $f_{p5}$. The Fourier spectra for the Pixel-by-Pixel analysis of the combined data
continues to $f_{p10}$. The
frequencies are
 listed in
Table \ref{pbpres2}.}\label{muldatacomp}
\end{center}
\end{figure*}
\end{center}

The cross-correlated line profiles of the three datasets were combined post cross-correlation to form a single dataset. The line profiles are
displayed in Figure
\ref{3lineprof}. Combining the line profiles instead of the individual spectra was the best way to preserve the signal-to-noise in the {\sevensize
MJUO} and La
Silla datasets which had
many more lines than the McDonald spectra. There was a small shift (6~km\,s$^{-1}$) in the systemic velocity between the McDonald and the other two
datasets. The McDonald spectra were shifted to ensure consistent data. The line profiles were scaled to have the same equivalent widths, a parameter
shown to remain constant by
the lack of variation in the
zeroth moment in the {\sevensize MJUO} data. The Fourier spectra of the {\sevensize MJUO} data alone and the combined datasets are compared in
Figure \ref{muldatacomp}.

The spectral windows for the {\sevensize MJUO} data alone and all three data sets are very similar, but, as expected, the addition of the multi-site
data reduces the
amplitude of the secondary peaks of the identified frequency. From Figure \ref{muldatacomp} it is clear that the
addition of the extra data increases the
amplitude of the Fourier peaks. This is an effect of not only the reinforced signal found in the other sites spectra, but also an
artificial amplification due to the increased base noise level. The frequency peaks
remain in nearly the same positions and show approximately the same repeating pattern. The one-day aliasing, although still present, has been reduced
to make the true peaks more evident. In general, the frequency identification becomes clearer with the addition of the multi-site data, despite it
being from small sets compared to the {\sevensize MJUO} data.

\vspace{5 mm}

\subsubsection{Pixel-by-Pixel (PbP)}\label{pbp2}

The line profiles showed many periodic variations in each pixel. The ten frequencies found in the combined dataset are
listed in Table \ref{pbpres2} and the Fourier spectra
displayed in Figure \ref{muldatacomp}. The first two frequencies found were the $f_2$ and $f_3$ from
\citet{2004AandA...415.1079A} from photometry and their $f_1$ shows up as our $f_{p4}$. Further frequencies $f_{p3}$ and $f_{p5}$ have credible three
bump variations in their line profiles. It is likely that $f_{p6}$ is a
one-day alias frequency
arising from the dominance of one single site ({\sevensize MJUO}  contributes about $80\%$ of the observations). The frequencies $f_{p7}$, $f_{p8}$
and $f_{p10}$ showed clear three
bump standard deviation profiles each with smooth phase changes. This increases the likelihood that they are real frequencies, although
the
possibility remains that they are from residual power from a previous poorly-defined frequency. Finally, $f_{p9}$ showed a clear four bump
standard
deviation profile at $3.14$~d$^{-1}$ and $4.14$~d$^{-1}$ frequencies when tested. As none of the previous frequencies showed a four-bump
pattern in the variation it is more likely this is a real frequency rather than a misidentification of an earlier frequency. 

A check on the
independence of the frequencies was done by prewhitening the spectra by 1-day aliases and 1-$f$ aliases of the first five identified frequencies.
Only $f_{p3} \pm 1$ had clear standard deviation profile variations resembling pulsation and none of the alternate identifications altered the
identification of following frequencies, indicating that they are all independent.

A second check of the variability of the line profile with different frequencies involves phasing the data to the proposed frequency and
examining the structure. To see the changes most clearly this was done using the residuals of the line profiles after subtraction of the mean. The
frequencies $f_{p1}$ to $f_{p5}$ in Figure \ref{tester} the residuals on the left side of each plot show the 'braided rope' structure typical of
non-radial pulsation, but for the other
frequencies this was less
clear. If we require pulsations
to produce regular changes in both red and blue halves of the line profile then we must discard $f_{p6}$, $f_{p7}$ and $f_{p9}$ as candidate
frequencies
leaving just
$f_{p8}$ and $f_{p10}$ to investigate further.

An investigation into the phase coverage of the data for each frequency can also be used as an indicator of the reliability of the frequency. All of
the
frequencies $f_{p1}-f_{p10}$ were well covered in phase space except $f_{p6}$, which was poorly sampled between phases 0.2-0.3 and 0.5-0.8 due to the
dominance of a single site in the data.

\begin{center}
\begin{table}\caption{Frequencies from the
Pixel-by-Pixel analysis of the combined dataset. The uncertainty estimate for the frequencies is $\pm0.0002$d$^{-1}$ as described in Section
\ref{freqres2}.
Frequencies with a strike-though were discarded as described in the text.}\label{pbpres2}
\begin{center}
\begin{tabular}{c c c c c c}
\\
ID & Freq.& Period & Amp. & Phase & Variation\\
 & (d$^{-1}$) & (d) &(rel. to $f_{p1}$) &  & Explained\\
 \hline
\\
 $f_{p1}$ & 1.3959 & 0.7164 & 1 & 0.286 & $13\%$\\
 $f_{p2}$ & 1.1863 & 0.8430 & 0.72 & 0.205 & $21\%$\\
 $f_{p3}$ & 1.6812 & 0.5948 & 0.93 & 0.267 & $30\%$\\
 $f_{p4}$ & 1.2157 & 0.8226 & 0.93 & 0.265 & $39\%$\\ 
 $f_{p5}$ & 1.5596 & 0.6412 & 0.80 & 0.230 & $45\%$\\
 $f_{p6}$ & \sout{1.0004} & 0.9996 & 0.55 & 0.158 & $48\%$\\
 $f_{p7}$ & \sout{1.2465} & 0.8022 & 0.50 & 0.143 & $50\%$\\
 $f_{p8}$ & 1.2743 & 0.7848 & 0.53 & 0.152 & $53\%$\\
 $f_{p9}$ & \sout{3.1392} & 0.3186 & 0.36 & 0.102 & $54\%$\\
 $f_{p10}$ & \sout{2.5803} & 0.3876 & 0.37 & 0.106 & $55\%$\\
\\
\end{tabular}
\end{center}
\end{table}
\end{center}

\subsubsection{Moment Analysis}\label{modeid}

The zeroth moment analysis shows few peaks that are similar to those from other methods. There is a cyclic variation around 1-cycle per day but with
the $1.0$~d$^{-1}$
peak missing due to the combination of multi-site data. The small amplitude of variations in the second moment indicates we have only small periodic
temperature variations to account for in the line profiles.

The first moment Fourier spectra showed clear peak frequencies similar to the PbP method. These are shown in the first section of Table \ref{momres2}.
The
amount of variation of the first moment was moderately well explained by the selection of the highest six peaks in the Fourier spectrum. Beyond
$f_{f6}$ the improvements to the explanation of variation are too small to be conclusive. When the PbP
frequencies were extracted from the first moment data (in approximately the same order as the highest
peak frequencies) a better fit to the variation was found. With the six frequencies, $f_{p1}-f_{p5}$ and $f_{p8}$, $76\%$ of the variation was
removed.
The frequencies found using {\sevensize SIGSPEC} closely matched the first four frequencies found in the highest peak method then reproduced PbP
frequencies
$f_{p3}$ and $f_{p7}$ as $f_{f9}$ and $f_{f10}$.

The Fourier spectra of the second moment also showed a few promising frequencies once the first peak at $0.0006$ d$^{-1}$ was removed.
The frequencies $f_{s2}$ to $f_{s4}$ match to, or match to one-day aliases of, frequencies found in the
PbP dataset. It is likely that $f_{s6}$ is a misidentification of $f_{p2}$. The second section of Table
\ref{momres2} catalogues the frequencies found. Both the highest peak and PbP frequencies accounted for around the same amount of variation as they
identified nearly all the same frequencies.

Frequency peaks found in the third moment are presented in the final section of Table \ref{momres2}. The
first few frequencies found using this method were generally similar to those found in the PbP method except the double
identification of $f_{t2}$. This is usually the result of small errors in the identification of a strong frequency, leaving residual power in the
Fourier spectrum. It is likely that this then impacts the following sequence
$f_{t5}$ to $f_{t10}$ which all seem close frequencies to other identifications, and possibly one-day aliases. The PbP frequencies identified
explained
more variation, accounting for a total of $83\%$ of the line profile variation. 

Overall the PbP variations have been shown to remove the variation from the data more accurately, demonstrated by the higher percentage of variation
removed for each moment. This leads us to conclude that the frequencies found using the highest peaks in the moment
methods are less reliable. It is notable that the {\sevensize SIGSPEC} identified frequencies are more reliable than the highest peak selection.

\begin{figure}
\centering
\subfigure{
\includegraphics[width=0.5\textwidth]{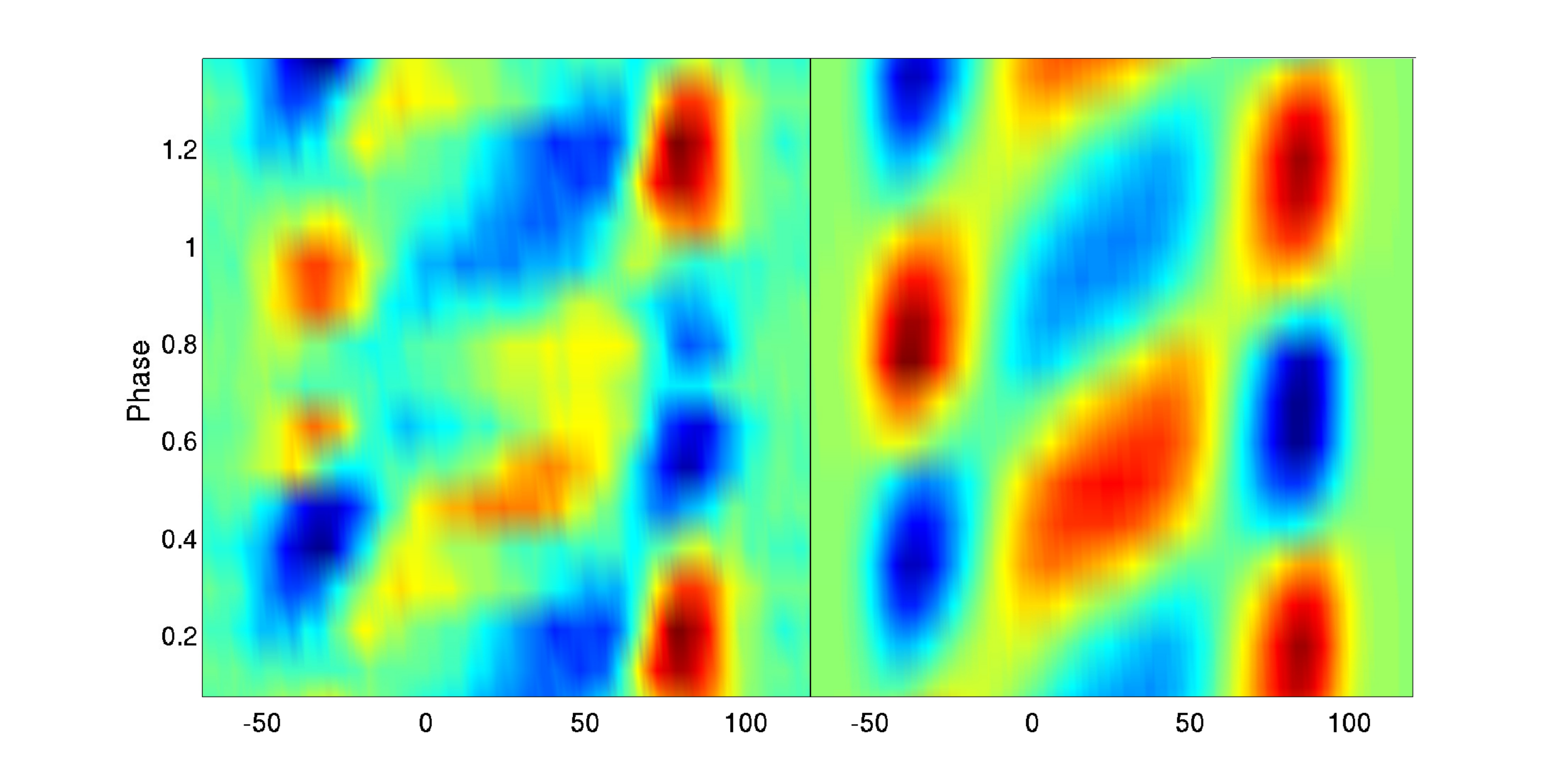}
\label{phaf12}
}
\subfigure{
\includegraphics[width=0.5\textwidth]{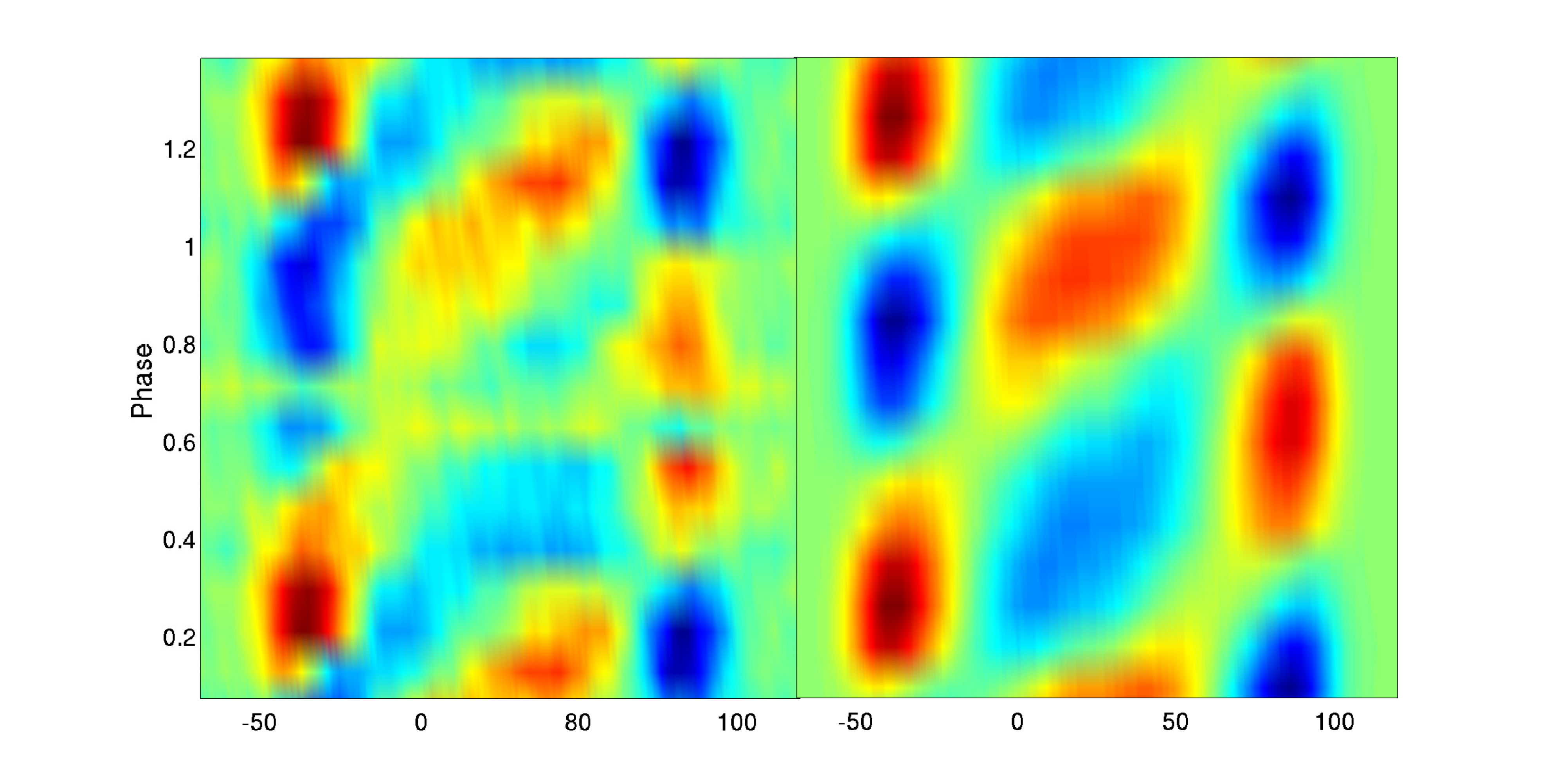}
\label{phaf13}
}
\subfigure{
\includegraphics[width=0.5\textwidth]{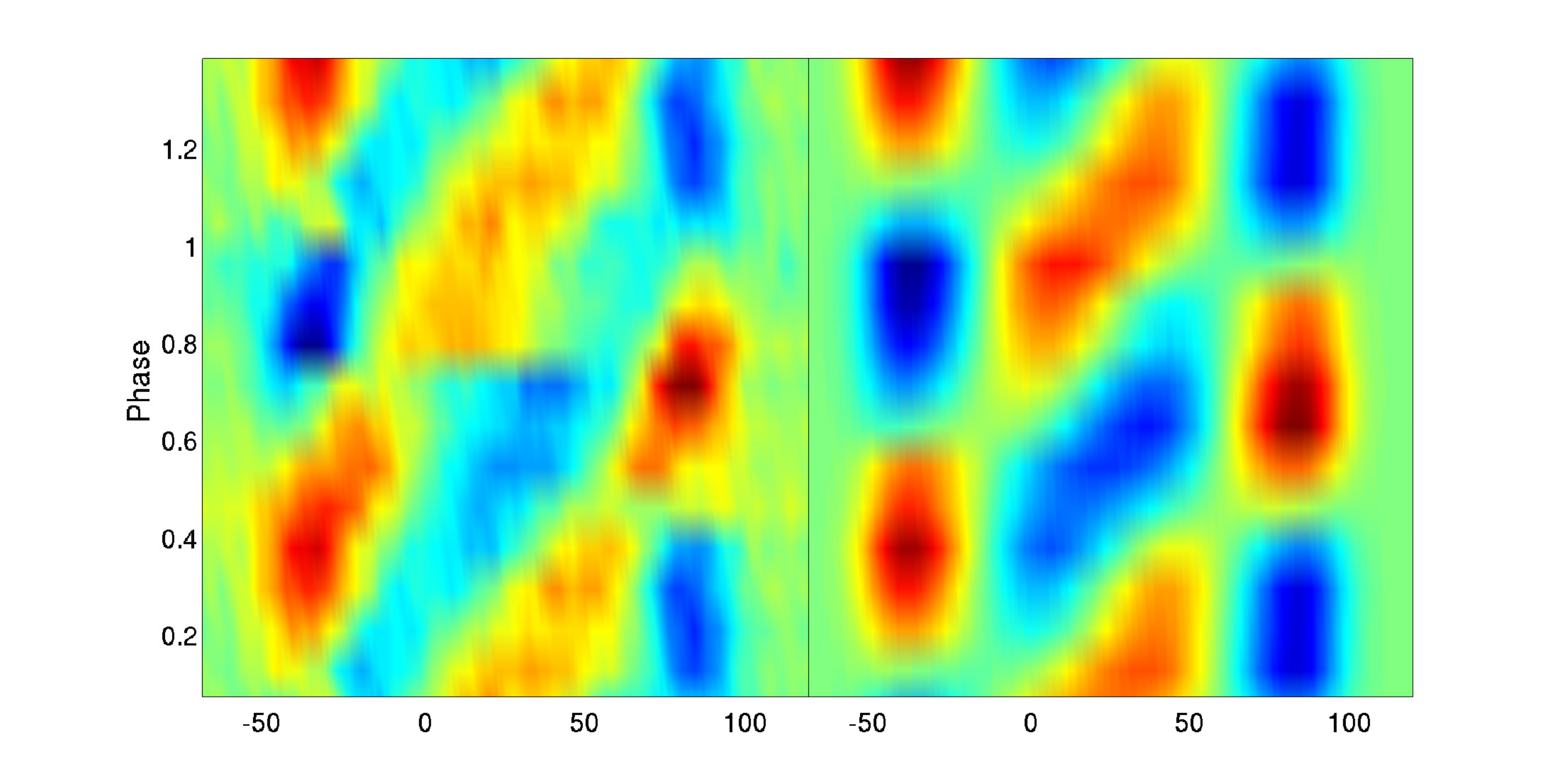}
\label{phaf14}
}
\subfigure{
\includegraphics[width=0.5\textwidth]{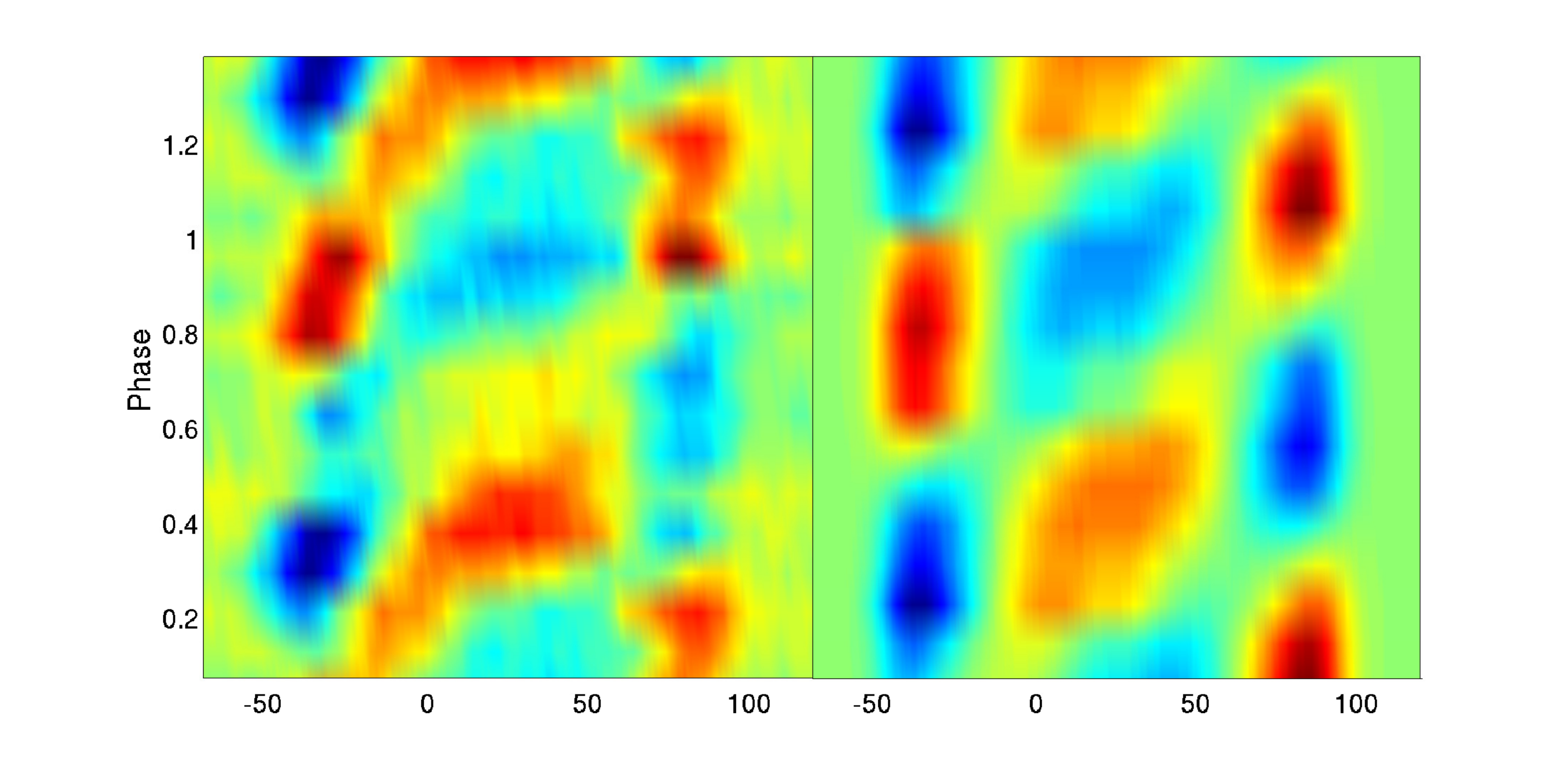}
\label{phaf15}
}
\subfigure{
\includegraphics[width=0.5\textwidth]{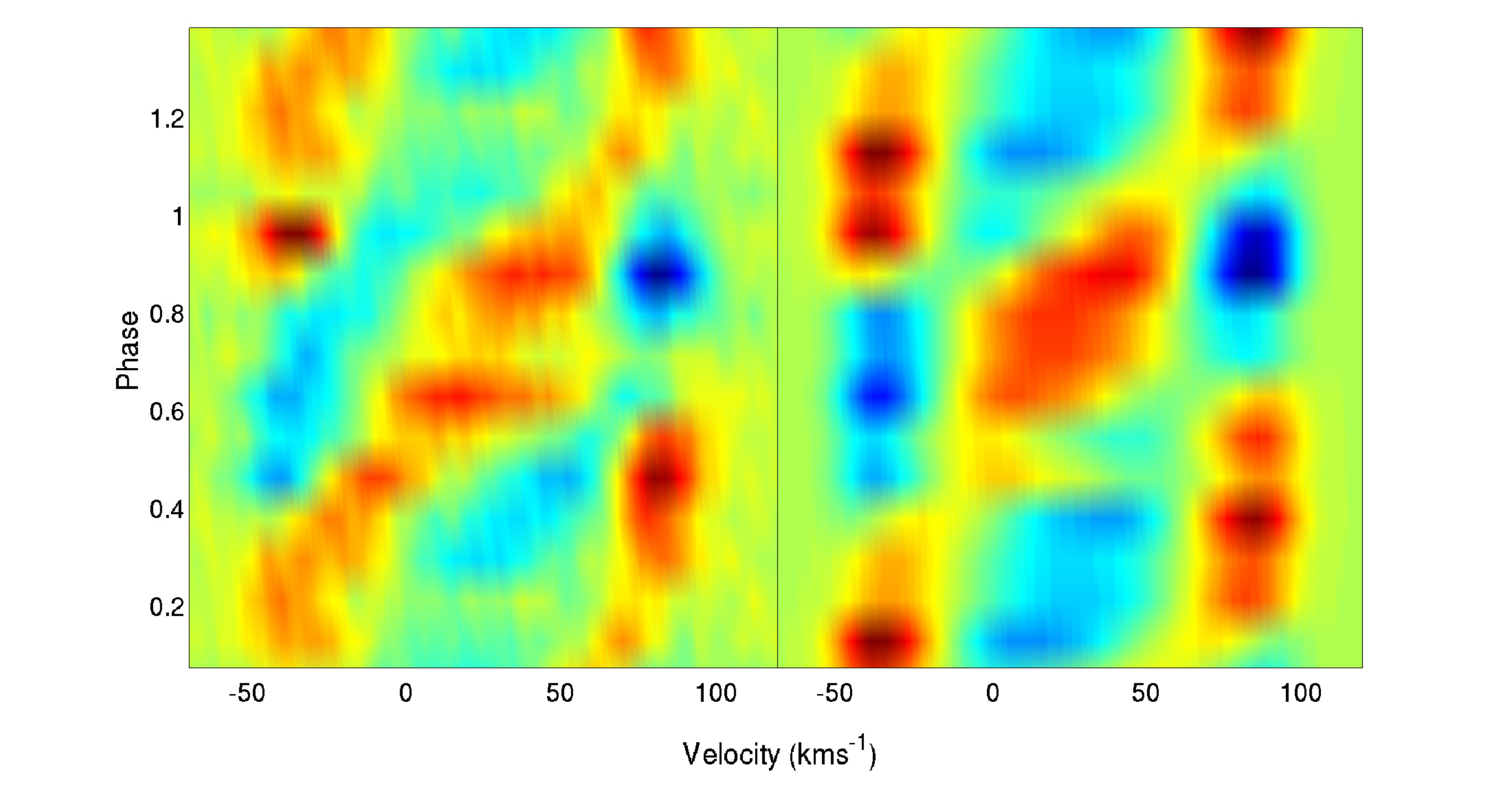}
\label{phaf16}
}
\caption{Line profile residuals phased on the frequencies identified $f_{p1}$ (top) to $f_{p5}$ (bottom) compared with synthetic profiles. The left
panel shows the observed profiles and the right is a synthetic model of the line profile using the modes identified in Section \ref{modeid2}. All are
(1,1) modes.}\label{tester}
\end{figure}

As an additional test of the robustness of the derived frequencies, a synthetic dataset, modelled on the time spacings and modes identified in Section
\ref{modeid2}, was created to compare with the PbP derived frequencies. The synthetic line profiles were phased and plotted to compare with
the real data in Figure \ref{tester}. The synthetic profiles strongly match the observed profiles, strengthening our confidence in the
frequency
identification.

\begin{center}
\begin{table}\caption{Resulting frequencies from the
First, Second and Third Moments found using peaks in the Fourier spectra and in {\sevensize SIGSPEC}}.\label{momres2}
\begin{center}
\begin{tabular}{c l l l l l }
\\
\multicolumn{5}{c}{First Moment}\\
\\
& \multicolumn{2}{c}{Highest Peak} & \multicolumn{2}{c}{{\sevensize SIGSPEC}} \\
\hline
ID & Freq. & Var. & Freq. & Sig.   \\
 \hline
 $f_{f1}$  & 1.1835  & $16\%$ & 1.18624 & 26 \\
 $f_{f2}$  & 1.6100  & $32\%$ &         &    \\
 $f_{f3}$  & 1.4017  & $42\%$ & 1.39591 & 26 \\
 $f_{f4}$  & 1.2247  & $52\%$ & 1.2156  & 28 \\ 
           & 1.2154  &        & 2.2273  & 21 \\
 $f_{f5}$  & 0.6490  & $57\%$ &         & 22 \\
 $f_{f6}$  & 1.5627  & $63\%$ & 1.5596  & 20 \\
 $f_{f7}$  & 0.7559  & $66\%$ &         &    \\
 $f_{f8}$  & 1.5179  & $69\%$ &         &    \\
 $f_{f9}$  &         &        & 1.6812  & 22 \\
 $f_{f10}$ &         &        & 0.2436  & 22 \\
 $f_{f11}$ &         &        & 2.1362  & 17 \\
 $f_{f12}$ &         &        & 0.0002  & 17 \\
 $f_{f13}$ &         &        & 2.7789  & 15 \\
 $f_{f14}$ &         &        & 0.6736  & 15 \\
\\
\multicolumn{5}{c}{Second Moment}\\
\\
&\multicolumn{2}{c}{Highest Peak} & \multicolumn{2}{c}{{\sevensize SIGSPEC}}\\
\hline
ID & Freq. & Var. & Freq. & Sig.\\
\hline
 $f_{s1}$ & 0.0006 & $7\%$  &        &    \\
 $f_{s2}$ & 0.2159 & $22\%$ &        &    \\
 $f_{s3}$ & 0.3767 & $38\%$ &        &    \\
          & 1.3962 &        & 1.3959 & 19 \\
 $f_{s4}$ & 1.5596 & $46\%$ & 1.5596 & 20 \\ 

 $f_{s6}$ & 2.1786 & $57\%$ &        &    \\
 $f_{s7}$ &        &        & 0.1622 & 25 \\
 $f_{s8}$ &        &        & 0.9449 & 23 \\
 $f_{s9}$ &        &        & 0.9061 & 19 \\
 $f_{s10}$ &        &        & 0.7382 & 19 \\ 
 $f_{s11}$ &        &        & 2.6809 & 17 \\
\\
\multicolumn{5}{c}{Third Moment}\\
\\
&\multicolumn{2}{c}{Highest Peak} & \multicolumn{2}{c}{{\sevensize SIGSPEC}}\\
\hline
ID & Freq. & Var. & Freq. & Sig.\\
 \hline
 $f_{t1}$  & 1.3987 & $17\%$ & 1.3956 & 31 \\
 $f_{t2}$  & 1.1864 & $40\%$ & 1.1862 & 25 \\
           & 2.1833 &        &        &    \\
 $f_{t3}$  & 1.6810 & $49\%$ & 1.6809 & 23 \\
 $f_{t4}$  & 1.3489 & $58\%$ &  &  \\
 $f_{t5}$  & 1.1661 & $64\%$ &  &  \\
 $f_{t6}$  & 2.5624 & $32\%$ & 0.5597 & 21 \\
 $f_{t7}$  & 2.7952 & $41\%$ &  &   \\
 $f_{t8}$  & 1.1716 & $49\%$ &  &   \\ 
 $f_{t9}$ & 3.5169 & $58\%$ &  &  \\
 $f_{t10}$ & 1.2658 & $64\%$ &  &  \\
 $f_{t11}$ &        &        & 1.2156 & 27 \\
 $f_{t12}$ &        &        & 2.1923 & 20 \\
 $f_{t13}$ &        &        & 0.2437 & 18 \\
\end{tabular}
\end{center}
\end{table}
\end{center}

\subsection{Frequency Results}\label{freqres2}

The frequencies chosen to proceed with in the analysis were those originally identified in the PbP results, $f_{p1}$ to $f_{p5}$ and $f_{p8}$
as noted in Table \ref{pbpres2}. This choice was made based on the prevalence of these frequencies through the other techniques used, the decreased
sensitivity of the PbP technique to asymmetric variations and generally observed higher signal-to-noise of the PbP frequencies.

It is clear from the above sections that, due to the inclusion of multi-site data, 1-day aliasing was not a big problem when identifying
frequencies present in these data.

Identifying the uncertainty range on the frequencies found is a complex task. When using {\sevensize SIGSPEC} we can use the
relation of \citet{2008AandA...481..571K}: \begin{equation}
\sigma(f) = \frac{1}{T*\sqrt{sig(a)}},
\end{equation} where T is the time base of observations in days and $sig(a)$ is the spectral significance of the frequency. For the frequencies
$f_{p1}$ to $f_{p10}$ identified in the first moment method this gives an uncertainty estimate of $\pm0.0002$. Additionally we used the
estimation from \citet{1999DSSN...13...28M} who propose 
 \begin{equation}
\sigma(f) = \sqrt{\frac{6}{N}} \frac{1}{\pi{T}} \frac{\sigma(m)}{A}.
\end{equation}
Here N is the number of observations. The value $\sigma(f)$ is a one-sigma uncertainty with an
amplitude (A) root-mean-square deviation $\sigma(m)$. This method was used to derive (an underestimate) of the uncertainties of the pixel-by-pixel
frequencies. The results for the frequencies $f_{p1}$ to $f_{p5}$ ranged from $\pm0.00013$ to $\pm0.00018$. Given the observed differences between the
frequencies identified in the PbP and the moment methods we view $\pm0.0002$ to be a good estimate of the uncertainty. 

The data were also analysed for possible frequency combinations. It was found that $f_{p5}(1.559)+ f_{p3}(1.681) =3.240$d$^{-1}$ which is possibly the
same
as $f_{p7}$(1.246) and strengthens our earlier removal of this frequency. 

It is clear from the above sections that many frequencies extracted from the data are robust as they appear in multiple methods with
clear line profile variations. The first five frequencies in the PbP analysis appear in almost every analysis method.
The high signal-to-noise in the line profiles for these frequencies makes them suitable for mode identification. The frequency $f_{p8}$, although
showing evidence for being an independent frequency, showed a misshapen standard deviation profile, which meant there was not enough certainty in
this frequency identification to proceed with a mode identification. The final PbP frequency $f_{p10}$ showed no evidence of appearing in other
analysis methods and was rejected for mode identification.

Even with all these frequencies identified, variations in the data remain. The PbP method only removed $50\%$ of the variation and the moment methods
up to $80\%$. Some of the remaining variation is due to the noise in the data and possible slight misidentifications of the frequencies, but it is
likely that
there remain multiple unidentified frequencies below the detection threshold of the present data. The presence of these further frequencies is
expected for $\gamma$ Doradus stars as they have dozens
and sometimes even hundreds of frequencies identifiable in photometry. The large number of similar amplitude frequencies make this star
challenging to study. However already we are able to see more frequencies than previously identified in spectroscopy, which makes this a
promising star for further study.

\begin{center}
\begin{table}\caption{Frequencies found in the {\sevensize  HIPPARCOS} white-light photometry. The first three frequencies match well with $f_{p4}$,
$f_{p5}$ and $f_{p7}$.
}\label{hipres}
\begin{center}
\begin{tabular}{c c c}
\\
\hline
Frequency (d$^{-1}$) & Uncertainty & Significance \\
\hline
 1.2158  & $\pm$0.0003 & 7 \\
 1.5525  & $\pm$0.0003 & 6 \\
 2.2449  & $\pm$0.0003 & 6 \\
 2.5665  & $\pm$0.0004 & 5 \\
\hline
\end{tabular}
\end{center}
\end{table}
\end{center}

\begin{center}
\begin{table*}\caption[Frequencies found in the Geneva photometry]{Frequencies (d$^{-1}$) found in each of the Geneva photometry filters and
significances using {\sevensize SIGSPEC}.}\label{gensig}
\begin{center}
\begin{tabular}{c|cc|cc|cc|cc|cc|cc|cc}
 &\multicolumn{2}{c}{B}& \multicolumn{2}{c}{B1}& \multicolumn{2}{c}{B2} & \multicolumn{2}{c}{G} & \multicolumn{2}{c}{U} & \multicolumn{2}{c}{V}&
\multicolumn{2}{c}{V1}\\
 & f.  & sig.& f. & sig.& f. & sig.& f.  & sig.& f.  & sig.& f.  &
sig.& f.  & sig.\\
\hline
$f_1$ & 1.215 & 13 & 1.215 & 13 & 1.215 & 12 & 2.218 & 11 & 1.215 & 13 & 1.215 & 11 & 1.215 & 12\\
$f_2$ & 0.09 & 8 & 0.09 & 8 & 0.09 & 8 & 0.393 & 6 & 0.07 & 5 & 0.09 & 7 & 0.095 & 8\\
$f_3$ & 1.396 & 8 & 1.396 & 8 & 1.396 & 8 & 0.071 & 6 &  &  & 1.396 & 7 & 1.396 & 6\\
$f_4$ & 2.567 & 5 & 1.565 & 5 & 3.56 & 4 & 3.32 & 5 &  &  & 4.32 & 4 & 3.32 & 6\\
$f_5$ &  &  &  &  &  &  & 3.19 & 4 &  &  &  &  & 2.95 & 4\\
\end{tabular}
\end{center}
\end{table*}
\end{center}

\section{Photometric Frequency and Mode Identification}\label{phfreqID2}

Two photometric datasets were analysed for frequencies. The first was a time series of white light observations from the satellite
{\sevensize  HIPPARCOS} \citep{1997ESASP1200.....P}. Specifically these were taken in the H$_p$ filter. The data span a period of 1166 days
from November 1989 to
February 1993 during which 122 measurements were taken. The second dataset was Geneva photometry taken on the 0.7m Swiss telescope at La Silla over a
period of 25 years from 1973 to 1997. A total of 174 observations for each filter in this time. This dataset has been extensively analysed in
\citet{2004AandA...415.1079A}. We present our re-analysis to complement the spectroscopic results above.

\subsection{{\sevensize  HIPPARCOS} Photometric Frequencies}\label{hipfreq}

The Fourier analysis of the {\sevensize  HIPPARCOS} data measurements have a range of 0.09 H$_p$ magnitudes with an average uncertainty of 0.1 H$_p$
magnitudes. It was difficult to identify clear frequencies from the Fourier spectra, but the highest peak was at 1.2701 d$^{-1}$. Using {\sevensize
SIGSPEC} we were able to identify the frequencies and their significances as shown in
Table \ref{hipres}.

\subsection{Geneva Photometric Frequencies}\label{genfreq}

The seven filters of the Geneva photometry provide us with enough information to extract frequencies and identify the $l$ modes. Each of the seven
filters showed very similar variation and this was reflected in the individual Fourier spectra. Frequency peaks were observed at
1.218 d$^{-1}$, 1.396 d$^{-1}$ and 0.0907~d$^{-1}$ (or one-day aliases) in most filters. To formalise the frequency identification we used
{\sevensize SIGSPEC} to identify the frequencies and their significances as shown in Table \ref{gensig}. These frequencies match well to those
previously found in the same data as is discussed further in Section \ref{disc}.

\subsection{Geneva Photometry Modes}\label{genmod}

The identification of the $l$ value of the spectroscopic frequencies $f_{p1}$ to $f_{p5}$ was attempted using the seven filters in the Geneva
photometric data. The
identification was done using the amplitude ratio method in {\sevensize FAMIAS} with modes from $l=1$ to $l=3$ tested in all filters. All five
frequencies were found to be solely consistent with $l=1$
modes distinguished primarily by the first amplitude ratio. The results are largely identical to \citet{2004AandA...415.1079A}, Their Figure 6 shows
the unique fit to the $l=1$ mode. This method, although not as powerful as spectroscopic mode identification, gives us independent
support to the $l=1$ mode fits of the star.

\section{Mode Identification of Spectroscopic Line Profiles}\label{modeid2}

\begin{center}
\begin{figure}
\includegraphics[width=0.45\textwidth]{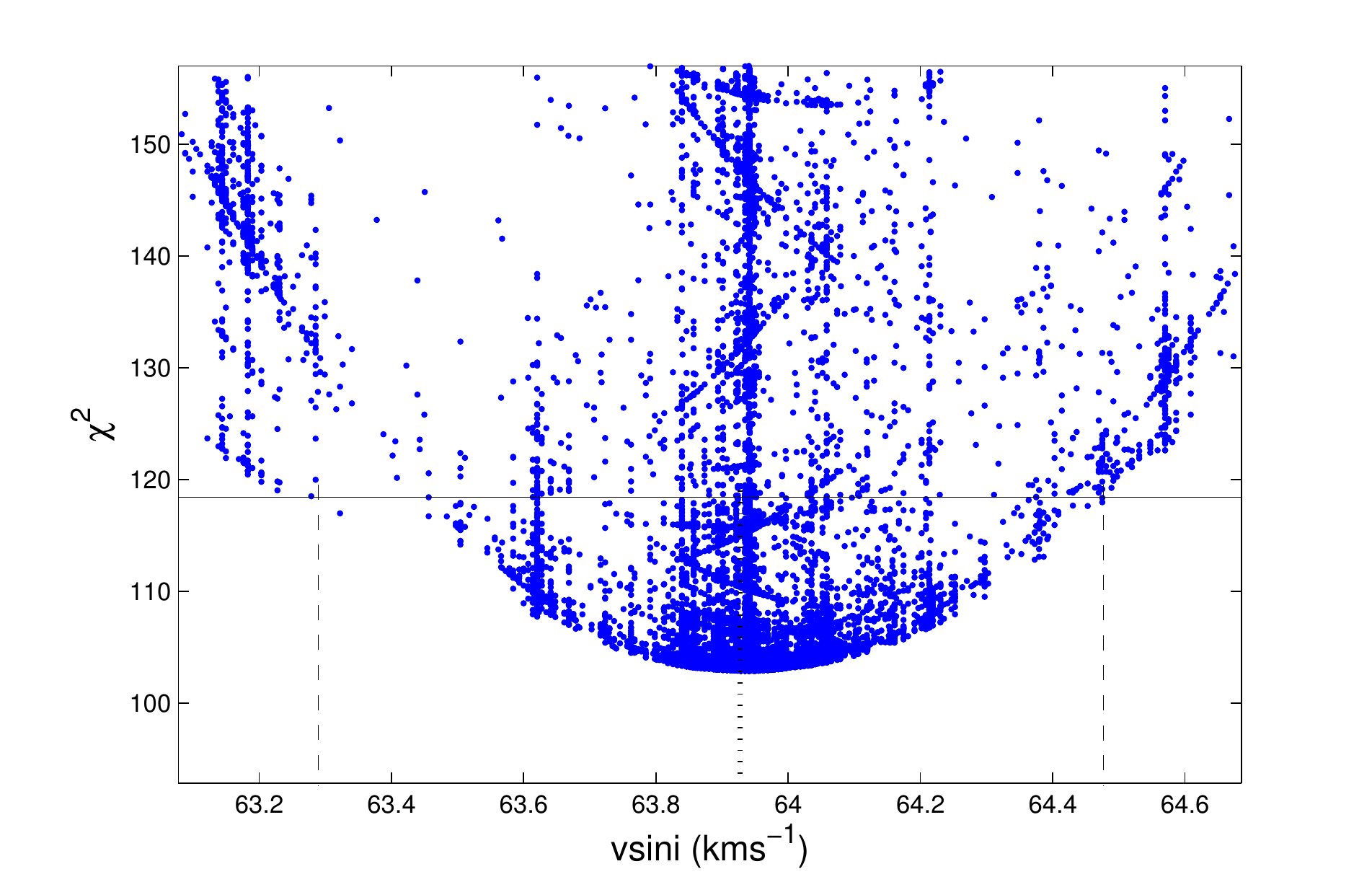}\caption{The $95\%$ confidence limit (solid line) of the zero-point fit to
the $v$sin$i$. The dashed lines show the limits at the confidence level and the dotted line indicates the minimum and hence best fit value.
}\label{inc_all}
\end{figure}
\end{center}

\begin{figure*}
\centering
\subfigure[Fit of the zero-point profile ($\chi^{2}=104$).]{  
\includegraphics[width=0.45\textwidth]{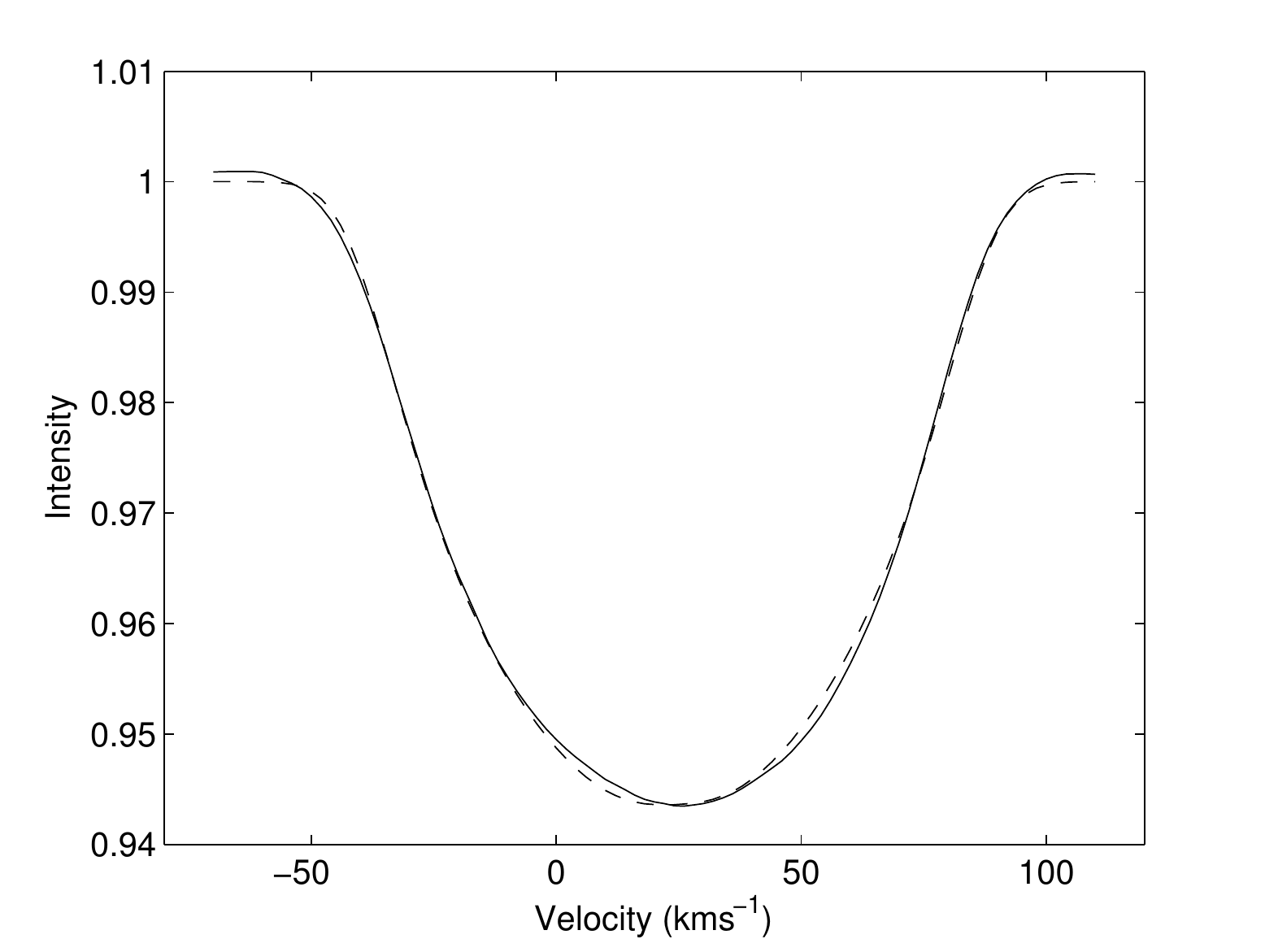}\label{zpfit}
}
\subfigure[Mode
identification of $f_{p1}=1.3959$~d$^{-1}$ ($\chi^{2}=7.7$).]{  
\includegraphics[width=0.45\textwidth]{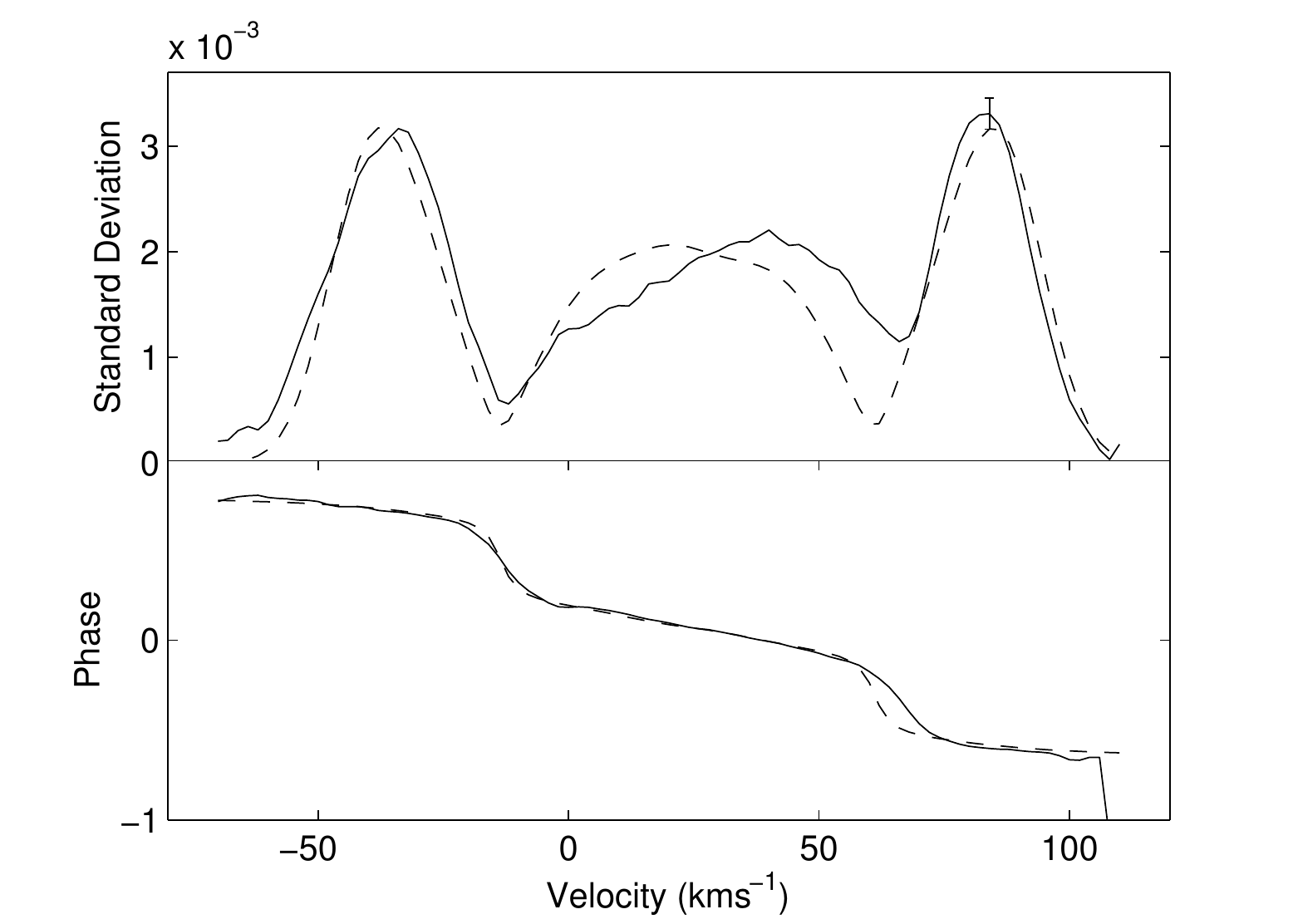}\label{mode2f1}
}
\subfigure[Mode
identification of $f_{p2}=1.1863$~d$^{-1}$ ($\chi^{2}=7.0$).]{  
\includegraphics[width=0.45\textwidth]{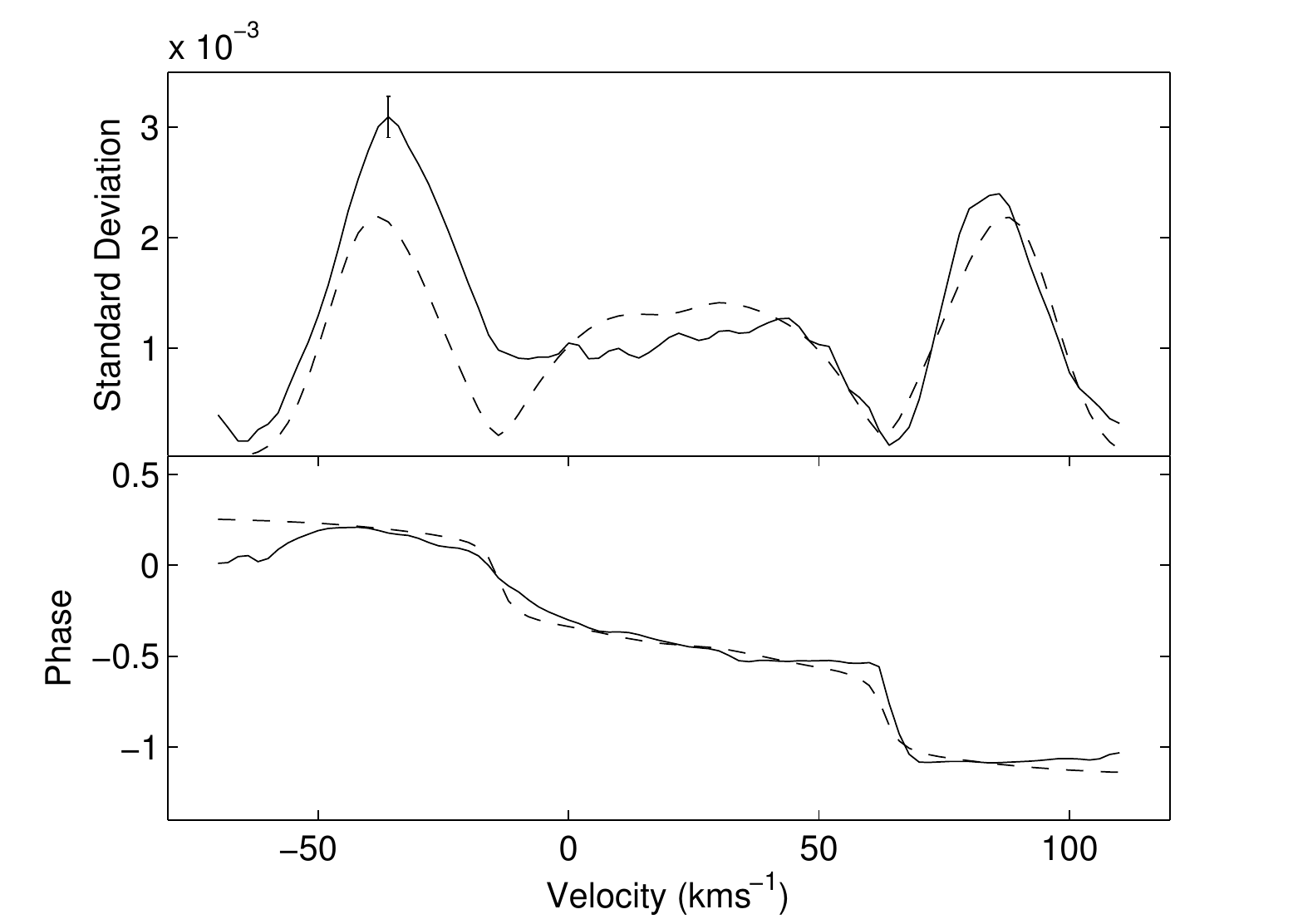}\label{mode2f2}
}
\subfigure[Mode
identification of $f_{p3}=1.6812$~d$^{-1}$ ($\chi^{2}=7.6$).]{  
\includegraphics[width=0.45\textwidth]{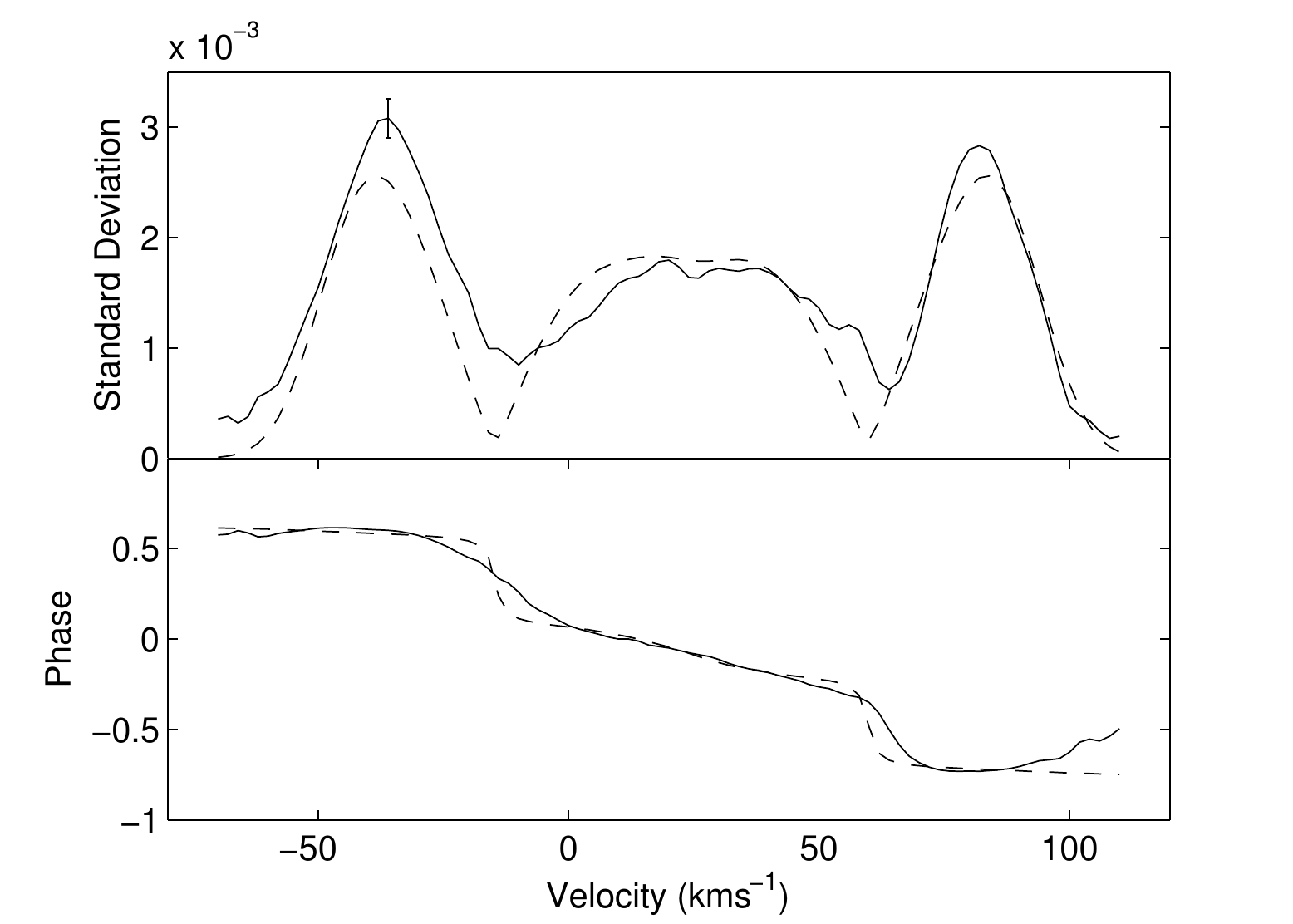}\label{mode2f3}
}
\subfigure[Mode
identification of $f_{p4}=1.2157$~d$^{-1}$ ($\chi^{2}=10.4$).]{  
\includegraphics[width=0.45\textwidth]{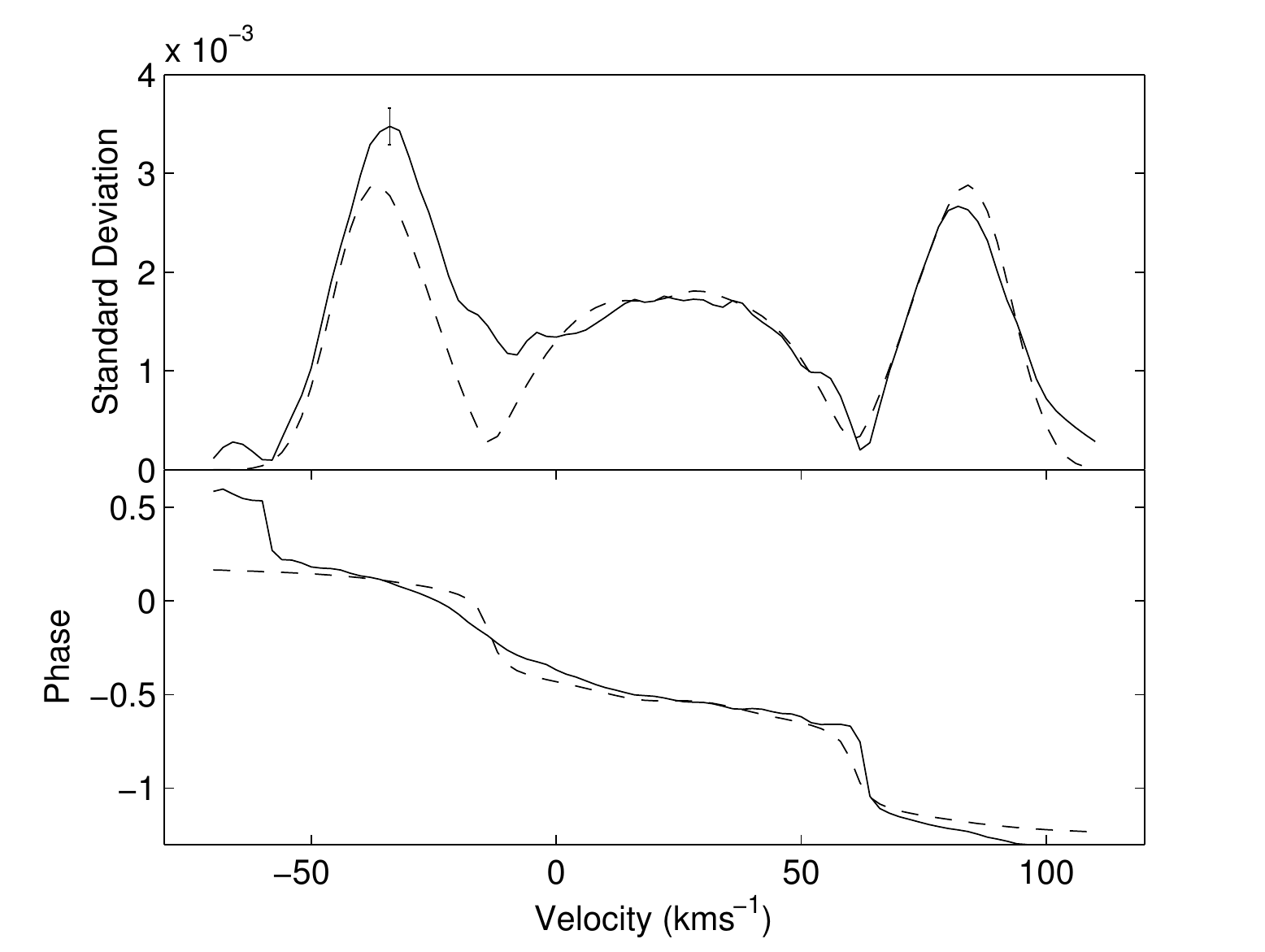}\label{mode2f4}
}
\subfigure[Mode
identification of $f_{p5}=1.5596$~d$^{-1}$ ($\chi^{2}=5.5$).]{  
\includegraphics[width=0.45\textwidth]{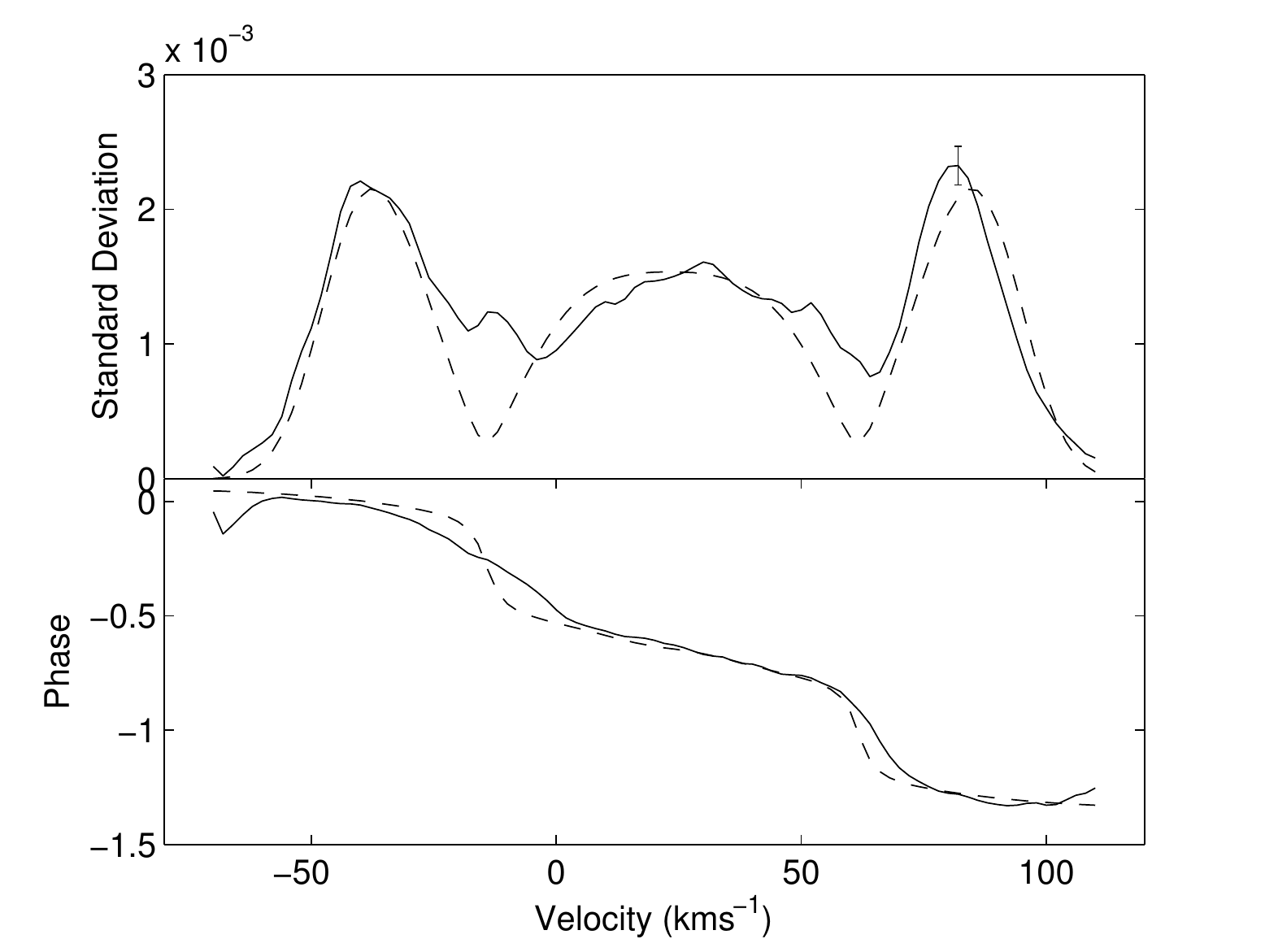}\label{mode2f5}
}
\caption[First Moment Fourier Spectra]{The fit
(dashed) of the mode identification to the mean line profile, variation and
phase (solid) of the five identified frequencies. All have been identified as (1,1) modes with the labelled $\chi^{2}$ for the best fit. An
indication of the maximum uncertainty is given on each plot.}
\end{figure*}

The mode identification was performed using the PbP frequencies $f_{p1}$ to $f_{p5}$, as chosen in Section \ref{freqres2} ($f_{p8}$ not
being considered due to the distorted line profile variations).
The modes of the individual frequencies were first identified and then a best fit, including all five frequencies, was computed.
Initially
the
zero-point profile was fitted to determine the basic line parameters. The best fit is shown in Figure \ref{zpfit}. The $v$sin$i$ was found to
converge at
$63.9$ km\,s$^{-1}$ with a $\chi^{2}$ of 103. Figure \ref{inc_all} shows the 95\% confidence limit on the determination of $v$sin$i$ which gives a
range
of $63.9 \pm0.5$ km\,s$^{-1}$. The zero-point fit set the initial line and stellar parameters of the space
searched. The details of the
parameters are given in Table
\ref{param2} and the resulting best fit modes are given in Table \ref{allmode2}. The unusually high values for mass and radius are discussed in
Section \ref{rotpul2}.

\begin{figure}
\centering
\subfigure[\protect$\chi^\protect{2\protect}\protect$ 
values for \protect$l\protect=0-3$ for
$f\protect_{p1}\protect=1.3959$d$\protect^\protect{-1\protect}$.\newline \hspace*{1.5em} Best fit value is for ($l,m$) = (1,1).]{  
\includegraphics[width=0.4\textwidth]{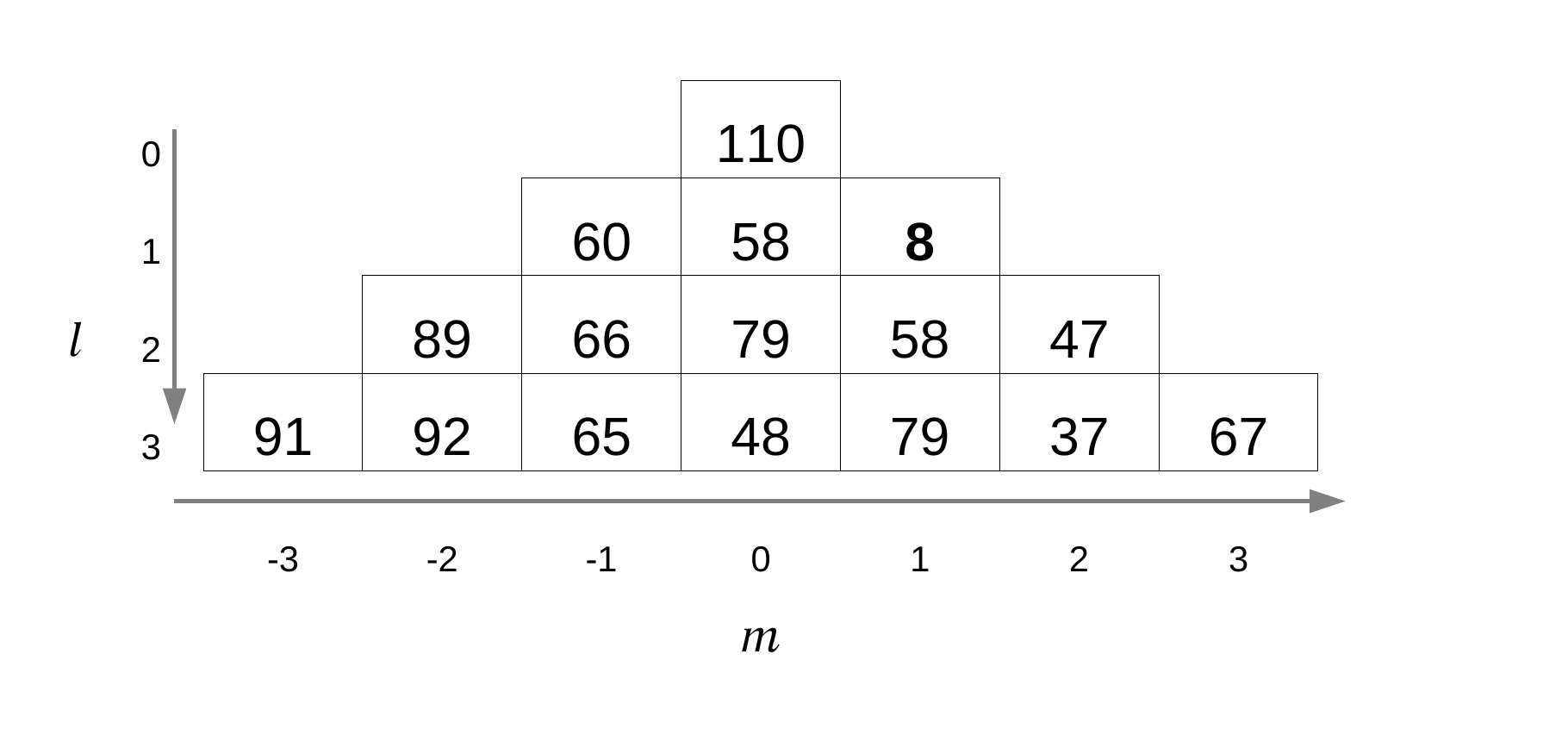}\label{best2fit1}
}
\subfigure[\protect$\chi^\protect{2\protect}\protect$ 
values for \protect$l\protect=0-3$ for
$f\protect_{p2}\protect=1.1863$d$\protect^\protect{-1\protect}$. \newline \hspace*{1.5em}
Best fit value is for ($l,m$) = (1,1).]{  
\includegraphics[width=0.4\textwidth]{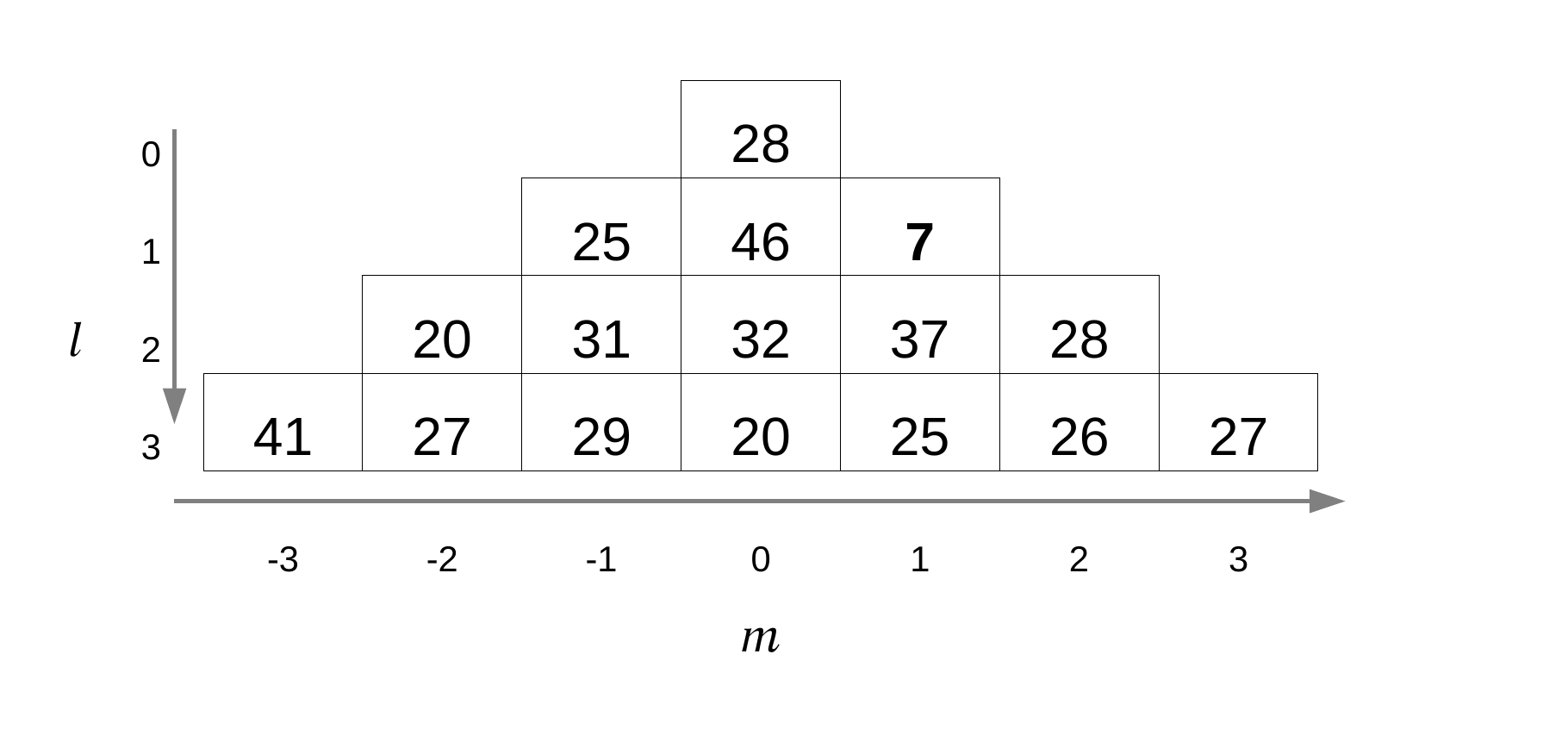}\label{best2fit2}
}
\subfigure[\protect$\chi^\protect{2\protect}\protect$ 
values for \protect$l\protect=0-4$ for
$f\protect_{p3}\protect=1.6812$d$\protect^\protect{-1\protect}$. \newline \hspace*{1.5em}
 Best fit value is for ($l,m$) = (1,1).]{  
\includegraphics[width=0.4\textwidth]{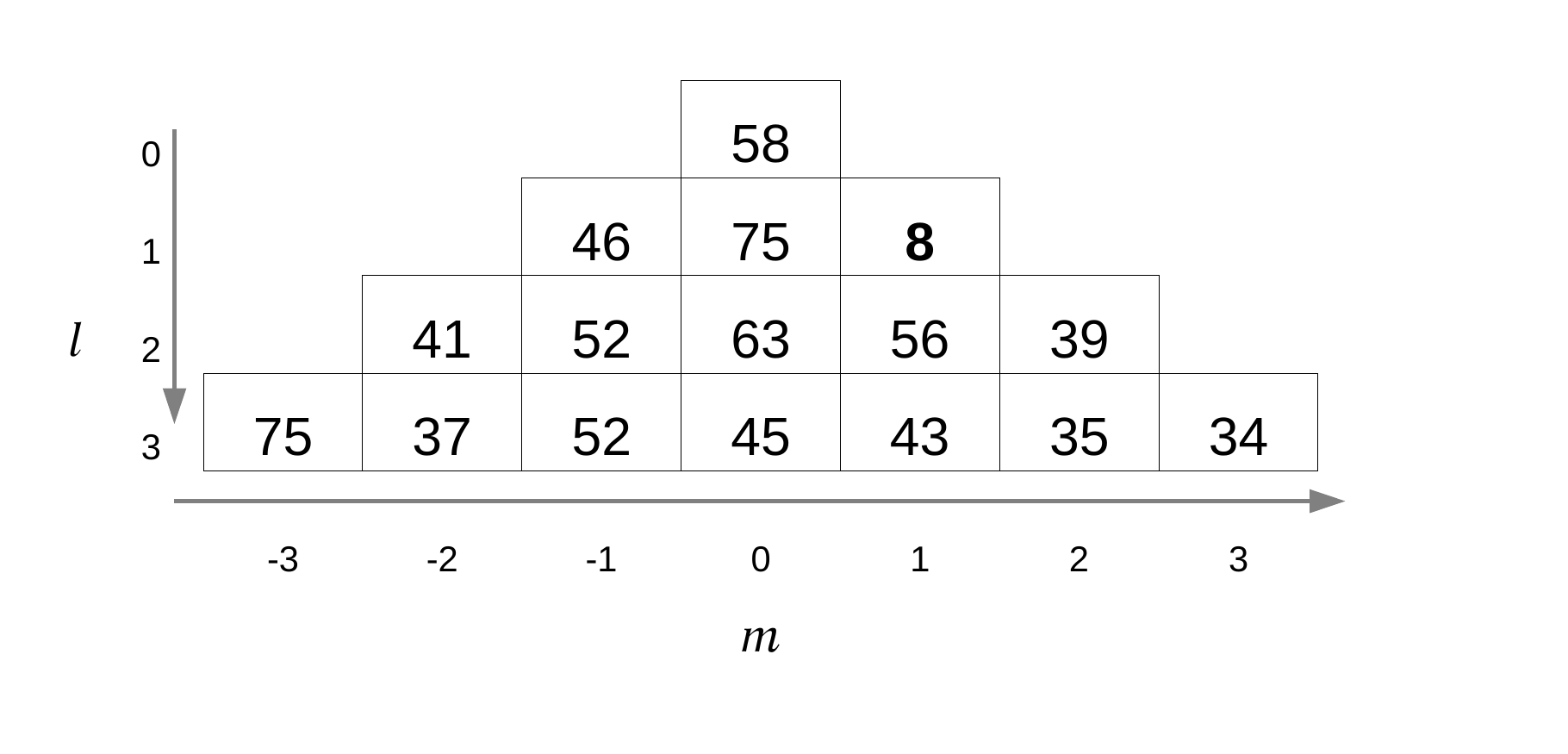}\label{best2fit3}
}
\subfigure[\protect$\chi^\protect{2\protect}\protect$ 
values for \protect$l\protect=0-3$ for $f\protect_{p4}\protect=1.2157$d$\protect^\protect{-1\protect}$. \newline \hspace*{1.5em}
Best fit value is for ($l,m$) = (1,1).]{  
\includegraphics[width=0.4\textwidth]{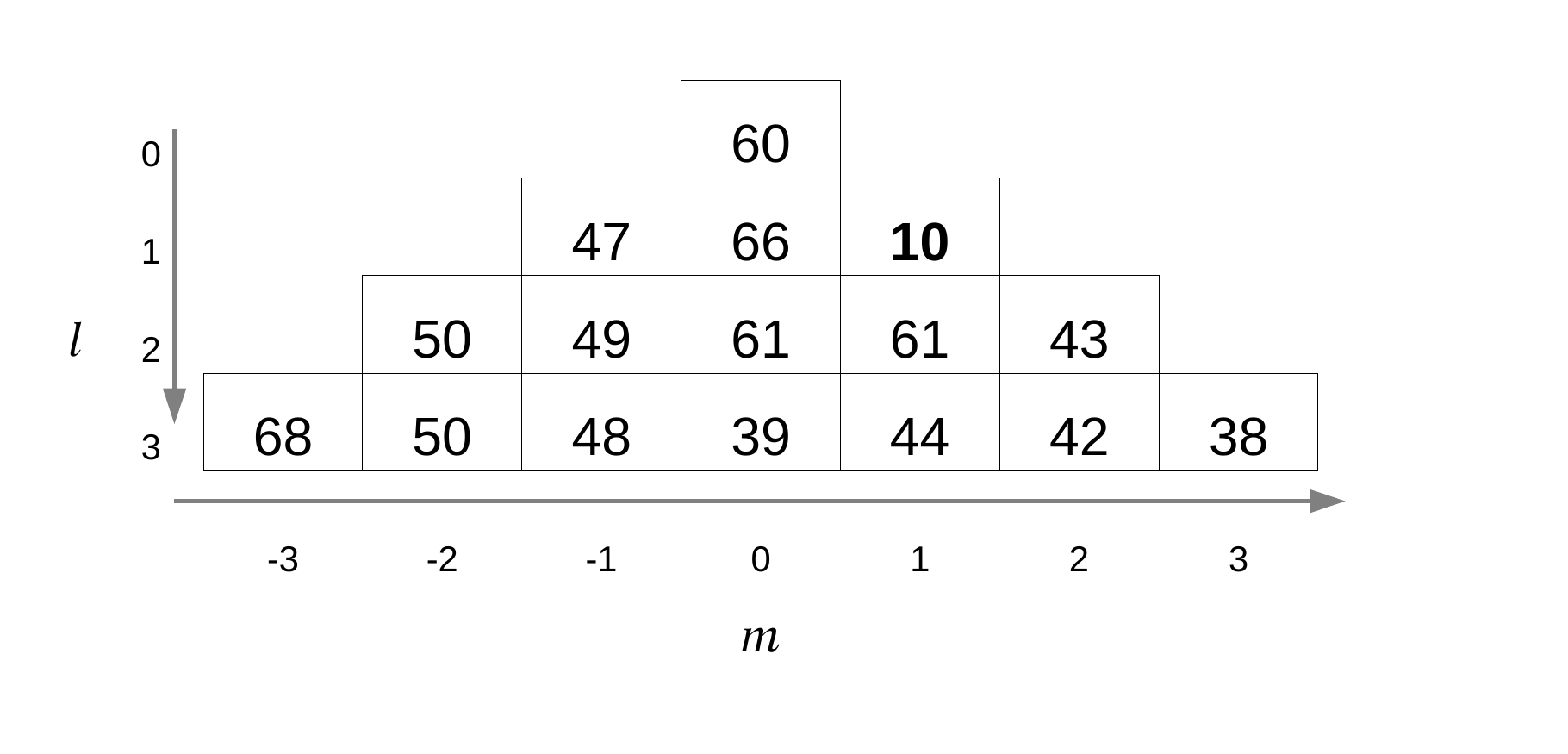}\label{best2fit4}
}
\subfigure[\protect$\chi^\protect{2\protect}\protect$ 
values for \protect$l\protect=0-3$ for $f\protect_{p5}\protect=1.5596$d$\protect^\protect{-1\protect}$. \newline \hspace*{1.5em}
Best fit value is for ($l,m$) = (1,1).]{
\includegraphics[width=0.4\textwidth]{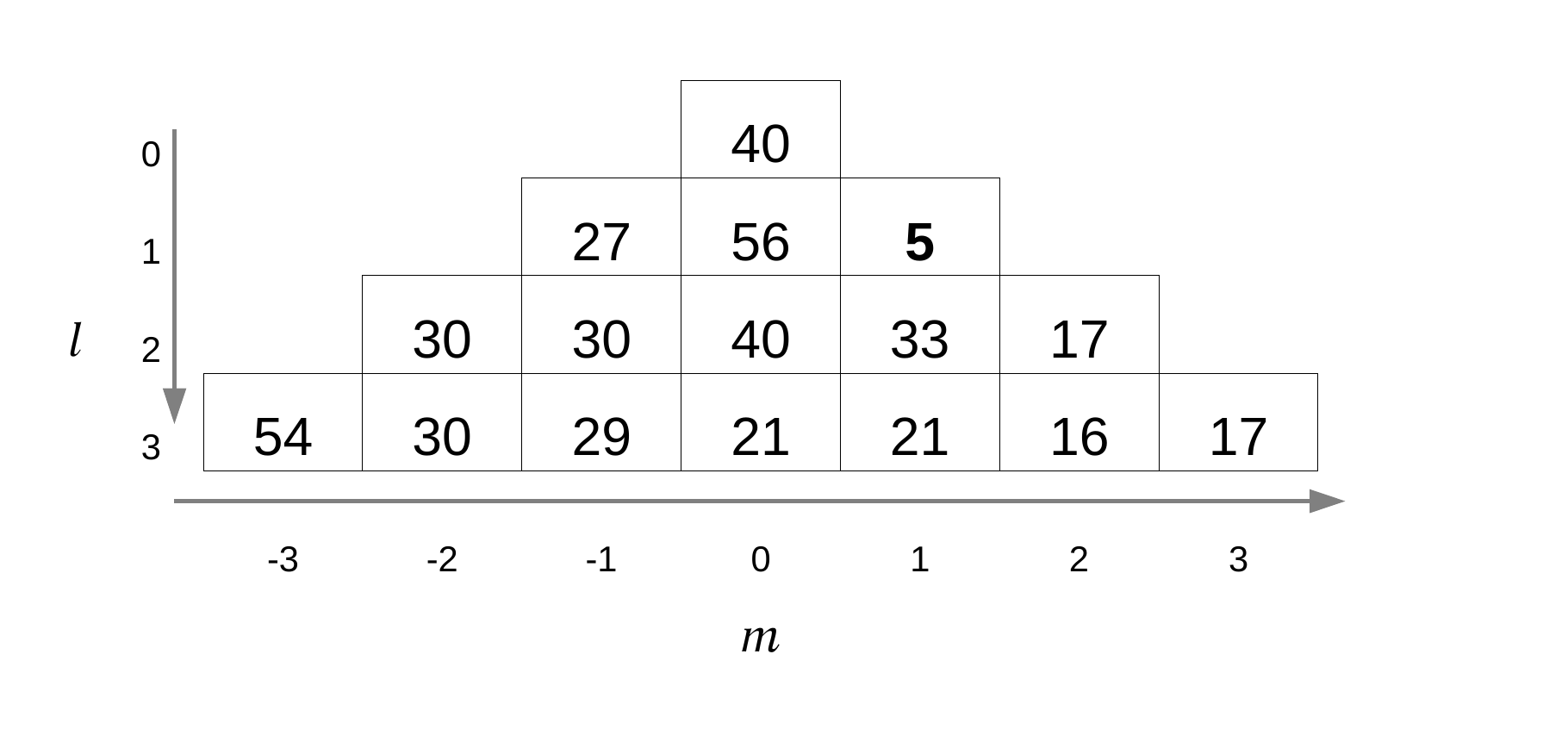}\label{best2fit5}
}
\caption{Lowest \protect$\chi^\protect{2\protect}\protect$ 
values for each possible (\protect$l,m\protect$) combination for the final identified frequencies. The best fit
\protect$\chi^\protect{2\protect}\protect$ is identified in bold.}
\end{figure}

The best fit models found for each individual frequency are plotted in Figures \ref{mode2f1} - \ref{mode2f5}. The results in Table \ref{allmode2} and
Figures \ref{best2fit1} to \ref{best2fit5} show that the mode identifications are well determined as there is no ambiguity in choosing the model with
the
lowest $\chi^{2}$. The inclination appears to lie in the region $20\degree-45\degree$, with the simultaneous fit giving a value of $27\degree$
for a  $\chi^{2}$ of 16.59. Formally the 95\% confidence limit of this parameter gives $27\degree$ $^{+52}_{-12}$, so it is poorly
constrained by the mode
identification. Given the $v$sin$i$ of the star, a value near $30\degree$ is plausible as described further in Section \ref{rotpul2}. The $v$sin$i$
value determined varies slightly between the
fits as it is modified to change the width of the standard deviation profile. The measured $v$sin$i$ of the star is best determined from the fit to
the zero-point profile given above.

The detection of five independent frequencies and the occurrence of a large number of (1,1) modes makes this star an excellent candidate for further
asteroseismic analysis. The prevalence of the (1,1) modes may indicate sequencing of the $n$-values, or the number of interior shells. It also is
possible that the (1,1) modes may be linked to the rapid rotation of the star. This is discussed in the context of all $\gamma$ Doradus stars in
Section \ref{disc}. 

\subsection{Rotation and Pulsation Parameters}\label{rotpul2}

The results of a preliminary mode identification were tested using the rotation and pulsation parameter tools in {\sevensize FAMIAS}. Using
mass = 1.5 $M_{\odot}$ and radius = 1.7 $R_{\odot}$ (typical for a $\gamma$ Doradus star), $v$sin$i$ = 64 km\,s$^{-1}$ and various inclinations
indicates the
rotational parameters of the star. Shown in Table \ref{paramrot} are the results for $i = 30 \degree$, which indicates that this star is not
approaching critical rotational velocity. This is the case for all tested values of $i$ in the range
$i = 10 \degree - 90 \degree$. The
rotational frequency of the star is dependent on the inclination, ranging from 4.28 d$^{-1}$ at $i = 10\degree$ to 0.74d$^{-1}$ at $i = 90\degree$.
The rotational frequency for a $\gamma$ Doradus star is expected to be on the same order as the pulsation frequency and these values fall within this
range.

\begin{center}
\begin{table}\caption{Rotational parameters and critical limits for a pulsating star with mass = 1.5 $M_{\odot}$,radius = 1.7 $R_{\odot}$,
$v$sin$i$ = 64 km\,s$^{-1}$ observed at an inclination of $i = 30 \degree$.}\label{paramrot}
\begin{center}
\begin{tabular}{ll}
\\
v$_{rot}$ & 128 km\,s$^{-1}$ \\
T$_{rot}$ & 0.67 d\\
$f_{rot}$ & 1.49 d$^{-1}$\\
v$_{crit}$ & 410 km\,s$^{-1}$\\
$v$sin$i_{crit}$ & 205 km\,s$^{-1}$\\
$i_{crit}$ & 9$\degree$\\
\end{tabular}
\end{center}
\end{table}
\end{center}

The pulsation parameter tool can be used to get an indication of the horizontal-to-vertical pulsation amplitude parameter ($\kappa$) and whether the
pulsation
and rotation frequencies fall within the mode determination limits of {\sevensize FAMIAS}. The $\kappa$ value falls between the values 50-800 for
frequencies
$f_1 - f_2$, which is above 1.0 as expected for $\gamma$ Doradus stars. These values then show the natural pulsational frequencies to lie between 0.2
d$^{-1}$ and 0.7 d$^{-1}$. Taking
the ratio $f_{rot}$/$f_{co-rot}$ gives values in the range 1-5 for the determined frequencies and modes (all $m = 1$). {\sevensize FAMIAS} has an
operational range of $f_{rot}$/$f_{co-rot} \leq 0.5$, being designed to deal with p-mode pulsations with much larger vertical pulsation components.
The very
high values of $\kappa$ for the g-mode pulsations in $\gamma$
Doradus stars and the high rotation rate of this star mean we are operating beyond this limit. To obtain a reasonable mode identification the mass and
radius were extended to non-physical values for a $\gamma$ Doradus star to provide observable amplitudes of pulsation (see Table \ref{param2}).
However the mass and radius are solely used for the calculation of the ratio $\kappa$, and do not affect other aspects of the fit. Making these
changes allowed us to
consider pulsations with $f_{rot}$/$f_{co-rot}$ values of 0.3-0.5 for the multi-frequency fit in Table \ref{allmode2}. 

Including more physics to more accurately
describe the effects of rotation of g-modes in current pulsation models would ultimately solve such problems. \citet{2003MNRAS.343..125T} investigated
the effect
of higher rotation on g-modes with high $n$ and low $l$ by increasing the Coriolis forces. It was seen that the identification of prograde modes
(defined here as $m > 0$, in \citealt{2003MNRAS.343..125T} as $m < 0$), have smaller differences in the rotating scenarios than retrograde modes.
Smaller values of $m$ are also less affected. There is also a discussion of this in \citet{2011ApJ...728L..20W} who estimate the true limits of the
$f_{rot}$/$f_{co-rot}$ ratio in {\sevensize FAMIAS} for various $m$ for a $\gamma$ Doradus range frequency and found {\sevensize FAMIAS} could
identify modes up to $f_{rot}$/$f_{co-rot} = 1$  for a $m = 1$ mode. These results indicate that the mode identification is unlikely to be affected by
the
rotation of this star. We can be confident that using non-physical values for mass and radius affects only the observed amplitude
of the
pulsation and not the mode, itself.

A convincing mode identification
of the
frequencies $f_1 - f_5$ was obtained. 

\begin{center}
\begin{table}\caption[Stellar parameters used in the mode
identification]{Stellar parameters used in the mode
identification. Fixed values for T$_{\rm eff}$, [M/H] and $\log g$ were taken from
\protect\citet{2008AandA...478..487B} and $v$sin$i$ values from the ranges in
\protect\citet{2008AandA...478..487B,2006AandA...449..281D,Dupret2005,2004AandA...415.1079A,2000AandA...361..201E}. }\label{param2}
\begin{center}
\begin{tabular}{lllll}
\hline
Parameter & Fixed & Min & Max & Step\\
 & Value &  &  & \\
\hline
Radius (solar units)  & × & 1 & 10 & 0.1\\
Mass (solar units) & × & 0.5 & 50 & 0.01\\
Temperature (K) & 7193 & × & × & ×\\
Metallicity [M/H] & 0.13 & × & × & ×\\
log $g$ & 4.18 & × & × & ×\\
Inclination ($\degree$) & × & 0 & 90 & 1\\
$v$sin$i$ (km\,s$^{-1}$) & × & 60 & 75 & 0.1\\
\hline
\end{tabular}
\end{center}
\end{table}
\end{center}

\begin{center}
\begin{table}\caption{Results of mode identification of all four
frequencies
individually after a least-squares fit is applied (lsf), and all four
frequencies simultaneously (sim).}\label{allmode2}
\begin{center}
\begin{tabular}{cccccc}
\hline
× & Mode  & $\chi^{2}$ & Inc.  & Amp  & Ph.\\
× & ID &  &  ($\degree$)  & (km\,s$^{-1}$)  & \\
\hline
$f_1$ lsf & (1,1) & 7.65 & 43.15  & 0.82 & 0.32\\
$f_2$ lsf & (1,1) & 7.00 & 35.79  & 0.74 & 0.80\\
$f_3$ lsf & (1,1) & 7.57 & 22.40  & 1.50 & 0.18\\
$f_4$ lsf & (1,1) & 10.43 & 33.11  & 1.50 & 0.72\\
$f_5$ lsf & (1,1) & 5.49 & 33.11 & 0.50 & 0.61\\
\hline
$f_1$ sim & (1,1) & 16.59 & 29.67 & 2.06 & 0.31\\
$f_2$ sim & (1,1)& & &  1.07 & 0.81\\
$f_3$ sim & (1,1) & & &  2.90 & 0.18\\
$f_4$ sim & (1,1) & & &  1.07 & 0.71\\
$f_5$ sim & (1,1) & & &  2.17 & 0.60\\
\hline
\end{tabular}
\end{center}
\end{table}
\end{center}

\section{Discussion}\label{disc}

This study allowed a direct comparison between single-site and multi-site data. From this we can judge the usefulness of large single-site datasets.
The findings show that although the addition of further sites increased the amplitudes of the frequencies, it also elevated the base noise level.
Additionally it is clear from the window function spectra in Figure~\ref{muldatacomp} that the 1-day aliasing pattern is reduced but not entirely
eliminated. The above leads us to conclude that the addition of multi-site data is useful, but not required to extract frequencies from sufficiently
large datasets. We must also require that the data from any one site have a sufficient number of observations to produce a balanced (well sampled in
phase space) mean line profile
in order to combine the cross-correlated line profiles with the highest precision.

The frequencies found in the PbP method were the most reliable and were consistent with all the other analysis methods. Past papers have
analysed the photometric data from {\sevensize HIPPARCOS} and also some multi-coloured photometry. In \citet{1999MNRAS.309L..19H}, the author finds a
frequencies of
2.18~d$^{-1}$ and 1.2155~d$^{-1}$ ($f_{p4}$).
In this paper we found frequencies $f_{p4}$, $f_{p5}$ and (2x) $f_{p8}$ in the {\sevensize HIPPARCOS} photometry. 

The multi-colour photometry was previously published by \citet{2004AandA...415.1079A}, who found frequencies equivalent to $f_{p4}$, $f_{p1}$ and a
1-day alias of $f_{p3}$. In the same data we find $f_{p4}$, $f_{p1}$ and $f_{p5}$. The mode identification showed the frequencies
$f_{p4}$, $f_{p1}$ and $f_{p3}$ to best fit $l=1$ modes, the same as found in this analysis. The same result was found by \citet{Dupret2005} who
showed that the three frequencies from \citet{2004AandA...415.1079A} are $l=1$ modes when using time-dependent convection models.

\citet{2004AandA...415.1079A} also describe the spectroscopic dataset taken with {\sevensize CORALIE} that was also used in this analysis.
The authors did not find
any
frequencies from the spectroscopic data alone (it is, as the authors note, too small a dataset for spectroscopic analysis). The imposition of the
$f_{p4}$, $f_{p1}$ frequencies found in their photometric analysis did provide some harmonic fits with low amplitudes.

A study applying the frequency ratio method to the frequencies found in \citet{2004AandA...415.1079A} was done by \citet{2005AandA...432..189M}, who
found three models consistent with $l=1$ modes. The models have a $T$= $6760K$, log$g=3.88-4.12$ and stellar ages around $2-3$ Myr.

Considering all the prior studies, all of the candidate frequencies we confirm ($f_{p1}$ to $f_{p5}$ and $f_{p8}$) are well supported. The mode
identification
is also partially confirmed by photometry. There is no suggestion of different frequencies observed in photometry and spectroscopy as there are from
some $\gamma$ Doradus stars (see \citealt{2011MNRAS.415.2977M,2008AandA...489.1213U}).

A prevalence of (1,1) modes in this star, and in $\gamma$ Doradus stars in general, is beginning to emerge. This includes two modes in
HD\,135825 \citep{2012MNRAS.422.3535B}, two in $\gamma$ Doradus \citep{1996MNRAS.281.1315B,Dupret2005}, one in HD\,40745 \citep{2011MNRAS.415.2977M},
one in HR8799 \citep{2011ApJ...728L..20W}, two in HD\,189631 \citep{mattttttttt} and two in HD\,65526
\citep{aaron}. This is possibly due to the large surface area covered in each segment of a (1,1) mode, meaning pulsations have larger amplitudes and
large changes in amplitude across the stellar surface. This indicates an observational selection effect. Additionally
\citet{2011MNRAS.415.3531B} discuss the prevalence of $m = \pm 1$ modes in a sample of $\gamma$ Doradus candidates from \textit{Kepler}
photometry. These authors assume the dominant frequency to be the rotational frequency and they then show further frequencies to be close to this
value. This
constrains the light maxima to  once per rotation cycle, requiring a $m = \pm 1$ mode. The occurrence of five of these such modes
in this star suggests some physical linking between them. The identification of several (1,1) modes in this star led to an investigation into the
period spacings of the six identified frequencies (Table~\ref{pbpres2}). The asymptotics of oscillation theory \citep{1980ApJS...43..469T} predicts a
characteristic period spacing for high-order g-modes of the same low degree ($l$) for sequential values of $n$. An investigation into the period
spacings of the PbP identified frequencies shows that the spacing between $f_{p1}$-$f_{p2}$ and $f_{p1}$-$f_{p3}$ to be close ($0.1266$d and $0.1215$d
respectively). This suggests they could be subsequent values of $n$ if we allow our frequencies to vary by $\pm 0.003$~d$^{-1}$. The frequencies
$f_{p4}$, $f_{p5}$ and $f_{p8}$ however do not fit with this spacing and additionally it may be that the close spacing of $f_{p2}$ and
$f_{p4}$ is inconsistent with current theoretical models. The identification of the frequencies could be further improved by photometric studies to
confirm this. Ultimately the sequencing of $n$-values could provide us with direct information about the stellar interior.

For the reasons above we propose this star as an excellent candidate with which to test asteroseismic models and potentially give us insight into the
pulsational behaviour of $\gamma$ Doradus stars.

\section{Acknowledgements}
This work was supported by the Marsden Fund administered by the Royal Society of New Zealand.

The authors acknowledge the assistance of staff at
Mt John University Observatory, a research station of the University of
Canterbury.

We appreciate the time allocated at other facilities for multi-site campaigns, particularly McDonald
Observatory and La Silla (European Southern Observatory).

Gratitude must be extended to the numerous observers who make acquisition of large datasets possible. We thank P. M. Kilmartin at MJUO and all the
observers at La Silla and the {\sevensize HIPPARCOS} team for their dedication to acquiring precise data.

This research has made use
of the {\sevensize SIMBAD} astronomical database operated at the CDS in
Strasbourg, France.

Mode identification results obtained with the software package {\sevensize FAMIAS} developed
in the framework of the FP6 European Coordination action {\sevensize HELAS}
(http://www.helas-eu.org/).

We thank our reviewer Gerald Handler for his helpful comments that improved this paper.

\section{Supporting Information}

Additional Supporting information may be found in the online version of this article:

Data file. The line profile data including Modified Julian Date, velocity (on a relative scale) and the intensity at each of the 120 velocity
sampling points for each profile. Line 1 is the axis scale, lines 2-478 are MJUO observations, lines 479-526 are La Silla observations and lines
527-586 are McDonald observations.

\label{lastpage}
\bibliography{references}{}
\bibliographystyle{mn2e}
\end{document}